\documentclass[11pt]{article}

\usepackage[T1]{fontenc}
\usepackage{lmodern}
\usepackage[width=17cm,height=23.5cm,centering]{geometry}
\usepackage{lipsum}
\usepackage{amsfonts}
\usepackage{graphicx}
\usepackage{epstopdf}
\usepackage{algorithmic}
\usepackage{amssymb}
\usepackage{mathtools}
\usepackage[bb=libus]{mathalpha}
\usepackage{bm,mathrsfs,xcolor,subfig}
\usepackage{verbatim}
\usepackage{hyperref}

\makeatletter
\newcommand*{\transpose}{%
	{\mathpalette\@transpose{}}%
}
\newcommand*{\@transpose}[2]{%
	\raisebox{\depth}{$\m@th#1\intercal$}%
}
\makeatother
\newcommand{\assign}{:=}
\newcommand{\mathi}{\mathrm{i}}

\newcommand{\tmem}[1]{{\em #1\/}}

\newtheorem{remark}{Remark}
\newcommand*{\email}[1]{\href{mailto:#1}{\nolinkurl{#1}} } 

\ifpdf
  \DeclareGraphicsExtensions{.eps,.pdf,.png,.jpg}
\else
  \DeclareGraphicsExtensions{.eps}
\fi
\graphicspath{{img/}}

\title{High-frequency homogenization for periodic dispersive media}

\author{
  Marie Touboul\thanks{Department of Mathematics, Imperial College London, London SW7 2AZ, UK (\email{m.touboul@imperial.ac.uk}).}
\and 
Benjamin Vial\thanks{Department of Mathematics, Imperial College London, London SW7 2AZ, UK (\email{b.vial@imperial.ac.uk}).}
\and 
Rapha{\"e}l Assier\thanks{Department of Mathematics, University of Manchester,
Manchester, UK (\email{raphael.assier@manchester.ac.uk }).}
\and 
S{\'e}bastien Guenneau\thanks{The Blackett Laboratory, Department of Physics, Imperial College London, London SW7 2AZ, UK (\email{s.guenneau@imperial.ac.uk}).}
\and 
Richard V. Craster\thanks{Department of Mathematics, 
UMI 2004 Abraham de Moivre-CNRS, 
Department of Mechanical Engineering, Imperial College London, London SW7 2AZ, UK. 
(\email{r.craster@imperial.ac.uk}). }
}

\usepackage{amsopn}
\DeclareMathOperator{\diag}{diag}

\newcommand{\beq}{\begin{equation}}
\newcommand{\eeq}{\end{equation}}
\newcommand{\ba}{\begin{eqnarray}}
\newcommand{\ea}{\end{eqnarray}}
\definecolor{colorcb}{rgb}{0 0.4470 0.7410}

\definecolor{colorcbq}{rgb}{0.4660 0.6740 0.1880}

\begin{document}

\maketitle

\begin{abstract}
  High-frequency homogenization is used to study dispersive media, containing inclusions placed periodically, for which the properties of the material depend on the frequency (Lorentz or Drude model with damping, for example). Effective properties are obtained near a given point of the dispersion diagram in frequency-wavenumber space. The asymptotic approximations of the dispersion diagrams, and the wavefields, so obtained are then cross-validated via detailed comparison with finite element method simulations in both one and two dimensions. 
  \end{abstract}


\section{Introduction}
Dispersive media have material properties that are frequency dependent and form an important and commonly occurring class of materials in physics, material science and engineering, they are of increasing importance in the design of advanced materials particularly in plasmonics \cite{Klimov2014,Maier2007}. In addition to practical relevance, dispersive media have interesting mathematical features for instance arising from resonances that lead to unusual material properties (e.g.\ artificial magnetism and negative refraction in metamaterials \cite{smith2004metamaterials}, zero-index band gaps \cite{Li2003}, or folded bands \cite{Chen2011}).\\
In electromagnetism, e.g.\ in optics, the electrodynamic properties of metals differ from those of dielectrics and, in particular, the relative permittivity is frequency dependent. The underlying physics is that free electrons form a plasma, are free to move, support a current, and create forces. At low frequencies in, say, microwaves, the effect is minimal, but as one approaches the optical regime, at TeraHertz (THz) frequencies, the plasma frequency becomes commensurate with the optical frequencies and the frequency dependence of the properties are important. Indeed the field has a long and distinguished history, the colours of metal glasses were investigated using the Drude model by Maxwell-Garnet \cite{MaxwellGarnet1904}, and the reflectance properties of metals and entire areas of physics such as plasmonics rely upon the properties captured within dispersive media. Historically, Drude, Lorentz and  Sommerfeld pioneered the material models that are now commonly used \cite{Levi2016}, and used in our numerical examples in section \ref{sec:numerical}. There is also broader usage of the Drude-Lorenz-Sommerfeld model in other fields as they are also well suited to describe resonant media, arising due to high material contrast of inclusions either in elasticity \cite{Auriault85,Auriault2012} or in electromagnetism \cite{Bouchitte2015,Felbacq2005}, or due to the presence of geometric resonances in acoustics e.g. Helmholtz resonators \cite{Fang2006,Haberman2016}. 

Structuring dispersive media to contain periodically arranged inclusions, or alternating layers \cite{Li2003} creates additional difficulties due to the introduction of geometric structure and the resulting scattering and reflection from inclusions and surfaces; there are applications for photonic crystals studied with Drude-like models \cite{moroz1999three}, creating complete electromagnetic band gaps,  or even for the more numerically challenging metallo-dielectric photonic crystals \cite{moroz2002metallo}. Critical to the study of wave propagation through periodic media is the concept of Bloch waves and of dispersion curves that connect frequency to the phase shift across a single cell of the periodic medium \cite{brillouin1946wave,Wilcox1978}; the resulting band diagrams then neatly encapsulate the essential physics of the structure, i.e. band-gaps of forbidden frequencies can arise where waves will not propagate in the structure. 
Our aim here is to provide homogenization models that allow us to side-step numerical issues, generate insight, and create effective media for wave propagation through dispersive media that contain periodically arranged inclusions, or consist of repeating layers. 

Classically, dynamic homogenization is understood as a low-frequency approximation to wave propagation in heterogeneous media. A particularly successful method, for periodic media, is the two-scale asymptotic expansion method and the notion of slow and fast variables  \cite{Bensoussan2011,Bakhvalov1989,cioranescu1999,Conca1997}. In the case of dispersive media, it has been extended to surfaces recently \cite{Lebbe2023}. The idea of high-frequency homogenization (HFH), introduced in \cite{Craster2010}, is to use similar asymptotic methods to approximate how the dispersion relation, and hence the media behave, near a given point in wavenumber-frequency space that satisfies the dispersion relation; recent work by {\cite{Guzina2019}} pushes this asymptotic analysis one order further. A uniform approximation is also derived, considering the fact that some branches do not intersect at the edges of the Brillouin zone but are close to each other. Other works concerned the inclusion of a source term \cite{Meng2020} and the derivation of the process in the time domain \cite{Harutyunyan2016}. The methodology has also been applied to several other configurations such as discrete lattice media \cite{Craster2010a,Colquitt2015}, frame structures \cite{Nolde2011}, optics \cite{Craster2011}, elastic plates \cite{Antonakakis2012}, full vector wave systems \cite{Antonakakis2014}, elastic composites \cite{Boutin2014}, reticulated structures \cite{Rallu2018}, or imperfect interfaces \cite{Assier2020,Guzina2021}. The two-scale approach is also connected  to homogenization near a neighbourhood of an edge gap in the context of approximations of operator resolvents \cite{birman2006homogenization} and there are connections into spectral theory. \\
Here we extend the HFH method to the case of dispersive media where the properties of the material depend on the frequency; this is not a routine extension as the dispersion curves are now complex and additional complications due to the frequency dependence, including resonances, now occur. In Section \ref{Sec:1D} we consider a one-dimensional (1D) setting of waves though a laminate of alternating layers, and high-frequency homogenization is applied for different cases: single eigenvalues at the edges, double eigenvalues at the edges, nearby eigenvalues at the edges, or single eigenvalues outside the edges (when no damping is considered); numerical examples are then presented to cross-validate the asymptotic approximations developed. In Section \ref{Sec:2D}, the asymptotic results are extended to two-dimensions (2D) and then cross-validated via comparisons with finite element simulations for metallic rods in a vacuum. The effective parameter obtained with HFH is also used to investigate properties of the material.  \\

\section{One-dimensional (1D) case}\label{Sec:1D}
\subsection{Setting}
We begin with the 1D case and consider linear waves propagating at a given angular frequency
$\omega$ through a dispersive periodic medium of periodicity $h (> 0)$ and with a
macroscopic characteristic length $L (> 0)$; physically this would correspond to a laminate of dispersive medium layers where the layers alternate with different material properties or a bimaterial string constructed from alternating dispersive media, see Fig. \ref{fig:3_scales} and for clarity of exposition and notation we will fix one of the media to be non-dispersive. We denote the physical space
variable as $X$; the material parameters $\hat{a}_h (X,\omega)$ and $\hat{ b}_h (X,\omega)$ are assumed to be $h$-periodic in $X$ and frequency dependent. The governing equation for the field $U_h$ is: 
\begin{equation}
    \label{Helmholtz_1D}
\frac{d}{dX}\left(\hat{a}_h(X,\omega)\frac{dU_h}{dX})\right)+\hat{ b}_h(X,\omega)\omega^2U_h=0.
\end{equation}
This equation is very general and the field $U_h$ describes the transverse electric field for s-polarisation, the transverse magnetic field for p-polarisation in electromagnetism, the displacement in elasticity, or the pressure in acoustics; the parameter $\hat{a}_h$ is then the inverse of the permeability, the inverse of the permittivity, the shear modulus, or the inverse of mass density, respectively, while $\hat{b}_h$ denotes the permittivity, the permeability, the mass density, or the compressibility, respectively. Henceforth we will assume the elastic setting in terms of notation.\\
The unit cell is divided into two parts distinguished by a ``volume fraction'' $\phi$. The left part is characterized by constant positive physical parameters, while in the right part they are frequency dependent and dispersive:  
\begin{equation}
\label{setting_1d_a}
\hat{a}_h(X,\omega)=\left\lbrace
    \begin{aligned}
   &{a}_0 &\text{ for } X\in(0,\phi h) \\
   &\hat{\mathbb{a}}(\omega) &\text{ for } X\in(\phi h,h) 
    \end{aligned}
    \right.
\end{equation}
\begin{equation}
\label{setting_1d_b}
\hat{ b}_h(X,\omega)=\left\lbrace
    \begin{aligned}
   &{ b}_0 &\text{ for } X\in(0,\phi h) \\
   &\hat{\mathbb{b}}(\omega) &\text{ for } X\in(\phi h,h).
   \end{aligned}
    \right.
\end{equation}
For the frequency dependence the example of a Lorentz type dependence is given in Appendix \ref{Sec:App1} and used in the numerical examples in section \ref{sec:numerical} .\\
 The edges of the periodic cell are assumed, without loss of generality, to be located at $X_n = nh$ for $n \in
\mathbb{Z}$, as illustrated in Figure \ref{fig:3_scales} (left). We
further assume that the interfaces across the edges of the periodic cells are perfect, implying continuity for the displacement $U_h$ and the stress $\hat{a}_h\frac{dU}{dX}$ there; the same is assumed within the unit cells at $n\phi h$. 

\subsection{Non dimensionalization}
\begin{figure}
    \centering
    \includegraphics[width=0.9\linewidth]{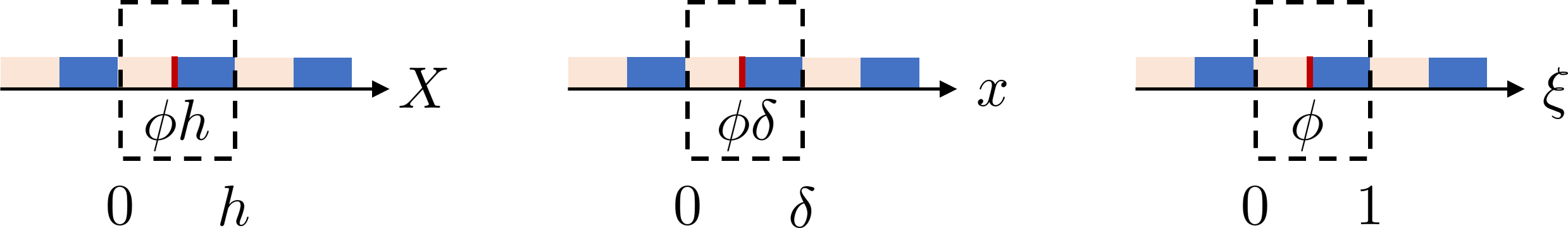}
    \caption{The three coordinate systems: (left) Physical coordinates, (middle) non-dimensional long-scale, (right) non-dimensional short-scale.}
    \label{fig:3_scales}
\end{figure}

To non-dimensionalize the physical problem, we introduce a reference wavespeed $c_0=\sqrt{a_0/ b_0}$ and the following non-dimensional quantities
\begin{equation}
  x = \frac{X}{L}, \quad \delta = \frac{h}{L}, \quad \Omega = \frac{\omega
  h}{c_0}, \quad \kappa = L k, \quad  u_{\delta} (x) = \frac{U_h (X)}{L} . 
  \label{eq:nondimparam}
\end{equation}
 Moreover, by periodicity
\begin{eqnarray*}
  \hat{ b}_h (X,\omega) = \hat{ b} \left( \frac{X}{h},\omega \right) & \quad \text{and} \qquad & \hat{a}_h (X,\omega) = \hat{a} \left( \frac{X}{h},\omega \right) ,
\end{eqnarray*}
where $\hat{ b}$ and $\hat{a}$ are 1-periodic in their first argument. These physical quantities are non-dimensionalized by introducing
\begin{eqnarray*}
     b \left( \frac{x}{\delta},\Omega
  \right) =
  \frac{\hat{ b}\left( \frac{X}{h},\omega \right)}{ b_0} & \quad \text{and} \qquad & a\left( \frac{x}{\delta},\Omega
  \right) = \frac{\hat{a}\left( \frac{X}{h},\omega \right)}{a_0}.
\end{eqnarray*}
Using these quantities, \eqref{Helmholtz_1D} is rewritten as the
non-dimensional governing equation
\begin{eqnarray}
  \delta^2 \frac{\mathrm{d}}{\mathrm{d} x} \left( a \left( \frac{x}{\delta},\Omega
  \right) \frac{\mathrm{d} u_{\delta}}{\mathrm{d}  x} (x) \right) + \Omega^2  b \left(
  \frac{x}{\delta},\Omega \right) u_{\delta} (x) & = & 0.
  \label{eq:goveqnondimproblem}
\end{eqnarray}
Upon introducing $$\mathbb{a}(\Omega)=\hat{\mathbb{a}}\left(\frac{c_0\Omega}{h}\right)\text{ and }\mathbb{b}(\Omega)=\hat{\mathbb{b}}\left(\frac{c_0\Omega}{h}\right),$$
the adimensionalized physical parameters depend only on the short scale $\xi=x/\delta$ and not on the long scale $x$ and become
\begin{equation}
\label{a_adim_setting}
a(\xi,\Omega)=\left\lbrace
    \begin{aligned}
   &1 &\text{ for } \xi\in(0,\phi) \\
   &\mathbb{a} (\Omega)&\text{ for } \xi\in(\phi,1) 
    \end{aligned}
    \right.
\end{equation}
\begin{equation}
 \label{rho_adim_setting}
 b(\xi,\Omega)=\left\lbrace
    \begin{aligned}
   &1&\text{ for } \xi\in(0,\phi ) \\
   &\mathbb{b} (\Omega) &\text{ for } \xi\in(\phi,1) .
    \end{aligned}
    \right.
\end{equation}
We still have continuity for $u_\delta$ and $a\frac{\mathrm{d}u_\delta}{\mathrm{d}x}$ at the points $n\delta$ and $n\phi\delta$ for $n\in\mathbb{Z}$ in the geometry setting of Figure \ref{fig:3_scales} (centre). \\

\subsection{Floquet-Bloch analysis}\label{Sec:BlochFloquet}
The periodicity of the parameters $a (\xi, \Omega)$ and $b (\xi,
\Omega)$ defined in (\ref{a_adim})-(\ref{rho_adim}), allows us to write the
solution of (\ref{eq:goveqnondimproblem}), as $u_{\delta} (x) =\mathfrak{u}_{\delta}
(x) e^{\mathi \kappa x}$, for a $\delta$-periodic function
$\mathfrak{u}_{\delta}$ and Bloch wavenumber $\kappa \in [0, \pi /
\delta]$. For any Bloch wavenumber $\kappa$, this implies that
$u_{\delta} \left( {\delta^+}  \right) = e^{\mathi \kappa \delta} u_{\delta}
(0^+)$ and $u_{\delta}' (\delta^+) = u_{\delta}' (0^+) e^{\mathi \kappa
\delta}$, where we use the prime symbol for differentiation. Using perfect contact conditions at $x=\delta$ and $x=\delta \phi$, the
whole problem is reduced to the unit cell $x \in (0, \delta)$ where
(\ref{eq:goveqnondimproblem}) must be satisfied together with
\begin{eqnarray}
  u_{\delta} (\delta^-) = e^{\mathi \kappa \delta} u_{\delta} (0^+) &
  \text{ and } & \mathbb{a} (\Omega) u_{\delta}' (\delta^-) = e^{\mathi \kappa
  \delta} u_{\delta}' (0^+) ,  \label{eq:Raph-BF-periodicity}
\end{eqnarray}
as well as
\begin{eqnarray}
  u_{\delta} (\delta \phi^-) = u_{\delta} (\delta \phi^+) & \text{ and } &
  u_{\delta}' (\delta \phi^-) =\mathbb{a} (\Omega) u_{\delta}' (\delta \phi^+).
  \label{eq:Raph-BF-perfect-contact}
\end{eqnarray}
The coefficients of (\ref{eq:goveqnondimproblem}) are piecewise constant (with respect to $\xi$) and so we have 
two different equations on $(0, \delta \phi)$ and $(\delta \phi,
\delta)$:

\begin{eqnarray}
  \left\{ \begin{array}{rcc}
    u_{\delta}'' (x) + \left( \frac{\Omega}{\delta} \right)^2 u_{\delta} (x) =
    0 & \text{for} & x \in (0, \delta \phi)\\
    u_{\delta}'' (x) + \left( \frac{\Omega}{\delta} \right)^2 \mathbb{g}
    (\Omega) u_{\delta} (x) = 0 & \text{for} & x \in (\delta \phi, \delta)
  \end{array} \right., & \text{ where } & \mathbb{g} (\Omega) \assign
  \frac{\mathbb{b} (\Omega)}{\mathbb{a} (\Omega)} \cdot
  \label{eq:Raph-BF-ODEs}
\end{eqnarray}
Equations (\ref{eq:Raph-BF-ODEs}) have solution
\begin{eqnarray}
  \left\{ \begin{array}{lc}
    u_{\delta} (x) =  \mathfrak{C}_1 \cos \left( \frac{\Omega x}{\delta}
    \right) +\mathfrak{C}_2 \sin \left( \frac{\Omega x}{\delta} \right) &
    \text{ for $x \in (0, \delta \phi)$}\\
    u_{\delta} (x)  =  \mathfrak{C}_3 \cos \left( \frac{\Omega
    \sqrt{\mathbb{g} (\Omega)} x}{\delta} \right) +\mathfrak{C}_4 \sin \left(
    \frac{\Omega \sqrt{\mathbb{g} (\Omega)} x}{\delta} \right) & \text{ for $x
    \in (\delta \phi, \delta)$}
  \end{array} \right. \cdot&  & 
\end{eqnarray}
Using the periodicity and interface conditions (\ref{eq:Raph-BF-periodicity})
and (\ref{eq:Raph-BF-perfect-contact}), the integration constants
$\mathfrak{C}_i$ satisfy the following linear system:
\begin{eqnarray}
  \bm{\mathcal{M}} (\mathfrak{C}_1, \mathfrak{C}_2, \mathfrak{C}_3,
  \mathfrak{C}_4)^{\transpose} & = & (0, 0, 0, 0)^{\transpose},  \label{eq:matrix-equation}
\end{eqnarray}
where the $4 \times 4$ matrix $\bm{\mathcal{M}}$ is given by

\begin{align*}
    \hspace{-8pt}\left( \hspace{-6pt}\begin{array}{cccc}
    e^{\mathi \kappa \delta} & 0 & - \cos \left( \Omega \sqrt{\mathbb{g}
    (\Omega)} \right) & - \sin \left( \Omega \sqrt{\mathbb{g} (\Omega)}
    \right)\\
    0 & e^{\mathi \kappa \delta} & \mathbb{a} (\Omega) \sqrt{\mathbb{g}
    (\Omega)} \sin \left( \Omega \sqrt{\mathbb{g} (\Omega)} \right) &
    -\mathbb{a} (\Omega) \sqrt{\mathbb{g} (\Omega)} \cos \left( \Omega
    \sqrt{\mathbb{g} (\Omega)} \right)\\
    \cos (\Omega \phi) & \sin (\Omega \phi) & - \cos \left( \Omega
    \sqrt{\mathbb{g} (\Omega)} \phi \right) & - \sin \left( \Omega
    \sqrt{\mathbb{g} (\Omega)} \phi \right)\\
    - \sin (\Omega \phi) & \cos (\Omega \phi) & \mathbb{a} (\Omega)
    \sqrt{\mathbb{g} (\Omega)} \sin \left( \Omega \sqrt{\mathbb{g} (\Omega)}
    \phi \right) & -\mathbb{a} (\Omega) \sqrt{\mathbb{g} (\Omega)} \cos \left(
    \Omega \sqrt{\mathbb{g} (\Omega)} \phi \right)
  \end{array}\hspace{-6pt} \right).
\end{align*}

Note that to get the second and fourth lines of $\bm{\mathcal{M}}$, we divided
through by $\Omega / \delta$. The system (\ref{eq:matrix-equation}) has
non-trivial solutions only when $\bm{\mathcal{M}}$ is singular. Upon dividing
through by \ $e^{\mathi \kappa \delta} 2\mathbb{a} (\Omega) \sqrt{\mathbb{g}
(\Omega)}$, the equation $\det (\bm{\mathcal{M}}) = 0$ reduces to the dispersion relation $\text{Disp}(\Omega,\kappa)=0$, where
\begin{equation}
  \label{eq:Disp}
\begin{aligned}
  \text{Disp}(\Omega,\kappa)&=\cos (\kappa \delta) - \cos (\Omega \phi) \cos \left( \Omega
  \sqrt{\mathbb{g} (\Omega)} (\phi - 1) \right) \\ &- \mathbb{d} (\Omega) \sin
  (\Omega \phi) \sin \left( \Omega \sqrt{\mathbb{g} (\Omega)} (\phi - 1)
  \right), 
\end{aligned}
\end{equation}
where the function $\mathbb{d}$ has been defined by
\begin{eqnarray*}
  \mathbb{d} (\Omega) & \assign & \frac{1}{2} \left( \frac{1}{\mathbb{a}
  (\Omega) \sqrt{\mathbb{g} (\Omega)}} +\mathbb{a} (\Omega) \sqrt{\mathbb{g}
  (\Omega)} \right) .
\end{eqnarray*}
Dispersion relations can usefully be thought of as nonlinear eigenvalue problems and, in
this context, it is known that if we take any open connected domain in the
$\Omega$ complex plane on which the entries of $\bm{\mathcal{M}}$ are holomorphic,
then there will be a finite (possibly zero) number of isolated zeros of $\det
(\bm{\mathcal{M}})$ within this domain (see e.g. Theorem 2.1 in
{\cite{guttel_tisseur_2017}}). Therefore the same is true for the solution of the
dispersion relation. One should however be careful about domains that contain
points for which the entries of $\bm{\mathcal{M}}$ are singular (e.g.\ branch points
or poles). Given the form of $\bm{\mathcal{M}}$, these potentially problematic
points are values of $\Omega$ for which $\mathbb{g} (\Omega) = 0$ or
$\mathbb{g} (\Omega) = \infty$. For spectral properties of absorptive and dispersive photonic crystals, we refer the reader to \cite{combes2002spectral} and \cite{Engstroem2009}, respectively.  More details are given in the next two paragraphs for the case of the Lorentz model (see Appendix \ref{Sec:App1} for the expression of the physical parameters in this case) as it is representative of issues that arise.

\paragraph{Points for which $\mathbb{g} (\Omega) = 0$ for the Lorentz model}

Given that $\mathbb{g} (\Omega) =\mathbb{b} (\Omega) /\mathbb{a} (\Omega)$,
these are the points for which either $\mathbb{b} (\Omega) = 0$ or $\mathbb{a}
(\Omega) = \infty$. So they are points that are solutions of (cf equations \eqref{a_adim} and \eqref{rho_adim})
\begin{eqnarray*}
  \left( 1 - \sum_{p\geq 0} \frac{\Omega_{p, i}^2}{\Omega (\Omega + \mathi \gamma_{p,
  i}) - \Omega^2_{D, p, i}} \right) = 0 & \quad \text{ for $i = 1$ or $i = 2$}. & 
\end{eqnarray*}
If there is only one term in each sum, these points can be written down easily
explicitly, but otherwise for several terms in the sums they have to be found
numerically; finding these points is straightforward.

\paragraph{Points for which $\mathbb{g} (\Omega) = \infty$ for the Lorentz model}

Given that $\mathbb{g} (\Omega) =\mathbb{b} (\Omega) /\mathbb{a} (\Omega)$,
these are the points for which either $\mathbb{b} (\Omega) = \infty$ or
$\mathbb{a} (\Omega) = 0$ and are solutions of
\begin{eqnarray*}
  \left( 1 - \sum_{p\geq 0} \frac{\Omega_{p, i}^2}{\Omega (\Omega + i \gamma_{p, i}) -
  \Omega^2_{D, p, i}} \right) = \infty & \quad \text{ for $i = 1$ or $i = 2$}. & 
\end{eqnarray*}
The points are found explicitly by nullifying the
denominators of each term in the sums and are given
for $i = 1, 2$ and $p$ by $\Omega =\mathcal{R}_{p, i}^{\pm}$, where
\begin{eqnarray*}
  \mathcal{R}_{p, i}^{\pm} & = & \frac{- \mathi \gamma_{p, i} \pm \sqrt{4
  \Omega^2_{D, p, i} - \gamma_{p, i}^2}}{2} \cdot
\end{eqnarray*}
In a neighbourhood of these points the theorem mentioned above does not apply,
and some of these will be {\tmem{accumulation
points}}. In other words, if we take any open connected set containing one of
these points, it will contain infinitely many zeros of the dispersion relation;
 this phenomenom is illustrated in Figure \ref{fig:Sheng_det_b}. It is
interesting to note that these points are independent of the choice of Bloch
wavenumber $\kappa$. In the remaining parts of this paper, we will aim to
provide an asymptotic homogenised approximation to the dispersion diagram and the
corresponding wave field in the vicinity of an exact solution $(\Omega_0,\kappa)$ of the
dispersion relation; our method works for points
$\Omega_0$ that are not too close to an accumulation point. In the vicinity of accumulation points, another approach is required as some form of resonance
is expected \cite{Brule2016,Mansuripur2017,Jakobsen2018,Besbes2022}. 

\subsection{High-frequency homogenization}
\begin{figure}
    \centering
    \includegraphics[width=0.9\linewidth]{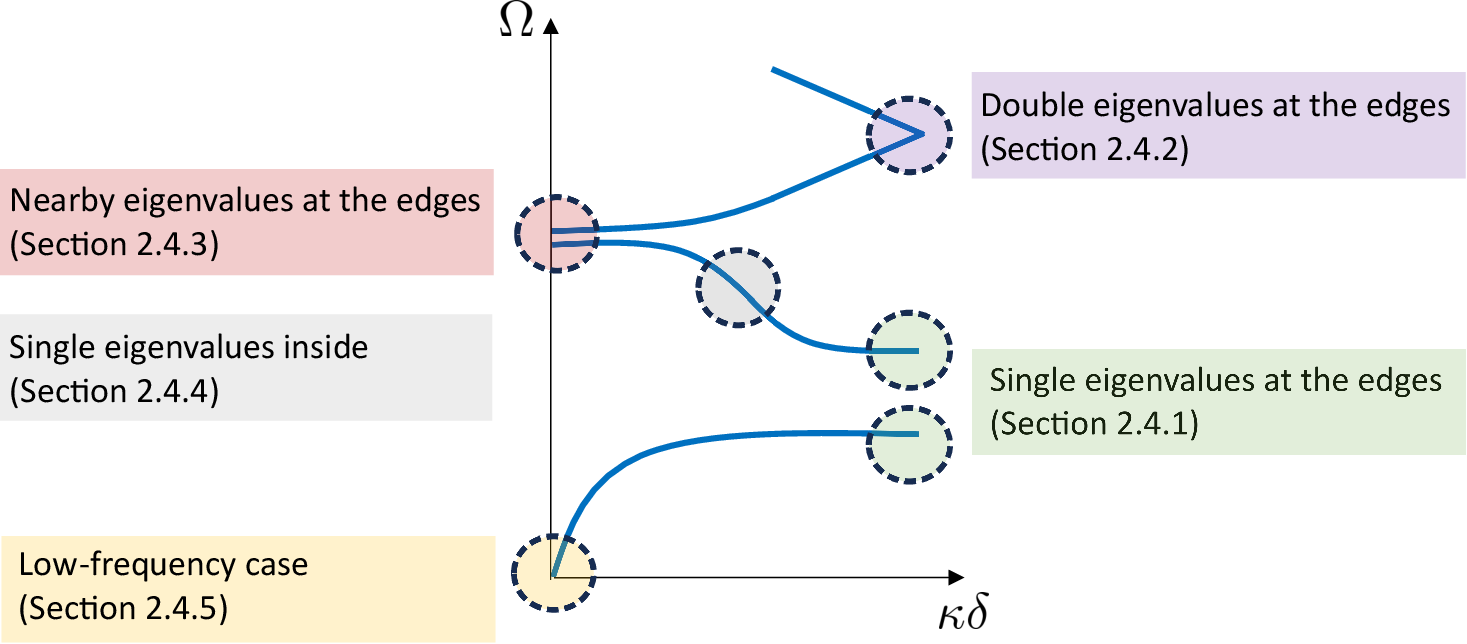}
    \caption{Different cases approximated with high-frequency homogenization.}
    \label{fig:diff_cases}
\end{figure}

We assume that $\delta \ll 1$ and we recall that $\xi=x/\delta$. To start with, we pick a frequency-wavenumber pair $(\Omega_0,\kappa)\in \mathbb{C}\times[0,\pi/\delta]$ that satisfies $\text{Disp}(\Omega_0,\kappa)=0$ and is such that we are not too close to an accumulation point. Following the two-scale expansion technique, we further assume the usual HFH ansatz for
the wave field $u_{\delta}$ and the reduced frequency $\Omega$:
\begin{eqnarray}
  u_{\delta} (x) = \sum_{j \geqslant 0} \delta^j u_j (x, \xi) & \text{ \ and \ } &
  \Omega^2 = \sum_{\ell \geqslant 0} \delta^{\ell} \Omega_{\ell}^2, 
  \label{eq:asymptoticrepfield}
\end{eqnarray}
where we treat $x$ and $\xi$ as two independent variables. The latter implies that  $\frac{\mathrm{d}}{\mathrm{d} x} \leftrightarrow
\frac{\partial}{\partial x} + \frac{1}{\delta} \frac{\partial}{\partial \xi}$. We will assume that 
\begin{equation}
    \label{eq:BF_cond}
    u_j (x, \xi + 1)=\mathrm{e}^{\mathrm{i}\kappa\delta} u_j (x, \xi),
\end{equation}
so we will restrict the analysis to (0,1) (see Figure \ref{fig:3_scales} right). \\
Using this ansatz, and considering that the physical parameters are piecewise constant, the governing equation \eqref{eq:goveqnondimproblem} becomes 
\begin{align}
  \sum_{j \geqslant 0} \left[ a(\xi,\Omega) \left\{  \delta^j  \frac{\partial^2u_j}{\partial \xi^2} +2 \delta^{j+1}\frac{\partial^2u_j}{\partial x\partial \xi}+\delta^{j+2}\frac{\partial^2u_j}{\partial x^2} \right\}  + \sum_{\ell \geqslant 0}
  \delta^{\ell + j} \Omega_{\ell}^2  b(\xi,\Omega) u_j \right]  =  0.
  \label{eq:allinonegoveq}
\end{align}
Importantly, $a$ and $ b$ depend implicitly on $\delta$ through $\Omega$ in the above expression. Therefore we have to write their expansion in powers of $\delta$. Up to the second order, we find that for $q =\{a, b\}$: 
\begin{equation}
    \label{expansion_coefficients}
    q(\xi,\Omega)=\mathcal{D}_0(q)+\delta \mathcal{D}_1(q)+\delta^2 \mathcal{D}_2(q)
\end{equation}    
with
\begin{eqnarray}
\label{eq:def_Di}
    &&\mathcal{D}_0(q) = q(\xi,\Omega_0) \label{eq:def_D0},\\
    &&\mathcal{D}_1(q) =  \frac{\Omega_1^2}{2 \Omega_0} \ \frac{\partial q}{\partial \Omega}(\xi,\Omega_0) \label{eq:def_D1},\\
    &&\mathcal{D}_2(q) = \frac{\Omega_1^4}{8 \Omega_0^2} \ \frac{\partial^2 q}{\partial \Omega^2}(\xi,\Omega_0) + \frac{1}{2\Omega_0} \left(\Omega_2^2 -  \frac{\Omega_1^4}{4\Omega_0} \right)  \ \frac{\partial q}{\partial \Omega}(\xi,\Omega_0).\label{eq:def_D2}
\end{eqnarray} 
We define $A_i = \mathcal{D}_i(a)$ and  $B_i= \mathcal{D}_i( b)$ for $i=\{0,1,2\}$, where we have to keep in mind that $A_i$ and $B_i$ depend on $\xi$ and $\left\lbrace\Omega_0,\cdots,\Omega_i\right\rbrace$. For the dispersion model we chose their expressions are detailed in Appendix \ref{Sec:App1}. \\
We also need to introduce  the average operator $\langle\cdot\rangle$ defined by
$$\langle g\rangle=\int_0^1 g(\xi)\mathrm{d}\xi$$ for any function $g$. \\
In the next sections, we will apply high-frequency homogenization to get asymptotic approximations of both the wavefields and the dispersion diagrams for all the relevant  cases which are represented in Figure \ref{fig:diff_cases}.

\subsubsection{Single eigenvalues at the edges of the Brillouin zone}\label{Sec:1D_simple}
 We first take points at the edges of the Brillouin zone, that is at $\kappa=0$ or $\kappa=\pi/\delta$ which is associated to periodic and antiperiodic conditions, respectively, for the fields. In these cases, we expect the dispersion relation to be locally quadratic i.e. we set $\Omega_1=0$. Indeed, the mapping $\Omega_n(k)$ denoting the dispersion diagram along a branch is holomorphic \cite{guttel_tisseur_2017} except around accumulation points and at singular points (when eigenvalues are no longer single ones); combined with reciprocity, i.e. $\Omega_n(k)=\Omega_n(-k)$, this gives that $\mathrm{d}\Omega_n/\mathrm{d}k=0$ at edges.  \\

\paragraph{Zeroth-order field}
Collecting the terms of order $\delta^0$, we get in $(0,1)$: 
\begin{equation}
    \label{syst_order0_main}
   A_0(\xi,\Omega_0)\frac{\partial^2u_0}{\partial\xi^2}(x,\xi)+\Omega_0^2B_0(\xi,\Omega_0)u_0(x,\xi) = 0
\end{equation}
where $A_0$ and $B_0$ are piecewise constants defined in \eqref{eq:def_D0}. We also have continuity for $u_0$ and $A_0\frac{\partial u_0}{\partial\xi}$ at $\phi$ and $0$ together with the 1-periodicity/antiperiodicity for $u_0$:
\begin{equation}
    \label{syst_order0_continuity}
    \left\lbrace
    \begin{aligned}
   u_0(\phi^-)&=u_0(\phi^+) \\
   \frac{\partial u_0}{\partial\xi}(\phi^-) &= \mathbb{a}(\Omega_0)\frac{\partial u_0}{\partial\xi}(\phi^+) \\
   u_0(1^-)&=\pm u_0(0^+) \\
   \mathbb{a}(\Omega_0)\frac{\partial u_0}{\partial\xi}(1^-)&=\pm\frac{\partial u_0}{\partial\xi}(0^+).
    \end{aligned}
    \right.
\end{equation}
As discussed in Section \ref{Sec:BlochFloquet}, we will build asymptotic approximations sufficiently far away from the accumulation points. We therefore know that there is a discrete set of eigenvalues and then choose $\Omega_0$ which is assumed to be a simple eigenvalue associated to the eigenfunction $\mathcal{U}_0(\xi,\Omega_0)$. The zeroth-order field is therefore
\begin{equation}
    \label{decompo_u0_single}
    u_0(x,\xi)=f_0(x)\mathcal{U}_0(\xi,\Omega_0),
\end{equation}
where the slowly varying amplitude $f_0(x)$ has to be determined. 
\paragraph{First-order field}
Collecting the terms of order $\delta$, and keeping in mind that $\Omega_1=0$, we get on $(0,1)$:
\begin{equation}
  \label{syst_order1_main}
A_0\frac{\partial}{\partial\xi}\left(\frac{\partial u_1}{\partial\xi}+2\frac{\partial u_0}{\partial x}\right)+\Omega_0^2 B_0 u_1=0
\end{equation}
together with periodicity/antiperiodicity for $u_1$, continuity for $u_1$ at $\phi$ and $0$, and continuity for $A_0\left( \frac{\partial u_0}{\partial x}+\frac{\partial u_1}{\partial\xi}\right)$ at $\phi$ and $0$.
Using \eqref{decompo_u0_single},  Equation\eqref{syst_order1_main} reduces to:
\begin{equation}
    \label{eq_main1_Omega1nul}
    A_0\frac{\partial^2u_1}{\partial\xi^2}+\Omega_0^2B_0u_1=-2A_0\mathcal{U}'_0(\xi)f'_0(x).
\end{equation}
Then, we write $u_1$ as: 
\begin{equation}
    \label{form_u1}
    u_1(x,\xi)=f_1(x)\mathcal{U}_0(\xi)+f_0'(x)V(\xi),
\end{equation}
where
\begin{equation}
\left\lbrace 
\begin{aligned}
    \label{system_h}
   & A_0 V''(\xi)+\Omega_0^2B_0V(\xi)=-2A_0\mathcal{U}'_0(\xi) \\
    &V \text{ is periodic/antiperiodic } \\
    &\text{$V$ and $A_0V'+A_0\mathcal{U}_0$ are continuous at $\phi$ and 0 }.
    \end{aligned}
\right.
\end{equation}
\paragraph{Second-order field}
Collecting terms of order $\delta^2$, we get on $(0,1)$:
\begin{equation}
    \label{syst_order2_main}
A_0\left( \frac{\partial^2u_0}{\partial x^2}+2\frac{\partial^2u_1}{\partial x\partial\xi}+\frac{\partial^2u_2}{\partial\xi^2}\right)+A_2\frac{\partial^2u_0}{\partial\xi^2}+\Omega_0^2(B_0u_2+B_2u_0)+\Omega_2^2B_0u_0=0,
\end{equation}
together with periodicity/antiperiodicity for $u_2$, continuity for $u_2$ at $\phi$ and $0$, and continuity for $A_0\left(\frac{\partial u_2}{\partial\xi}+ \frac{\partial u_1}{\partial x}\right)+A_2\frac{\partial u_0}{\partial\xi}$ at $\phi$ and $0$. Consider now the equation 
$$\langle {u_2}\times\eqref{syst_order0_main}-u_0\times{\eqref{syst_order2_main}}\rangle=0.$$
After integration by part, some algebra and dividing through by $f_0$ we get: 
\begin{equation}
    \label{syst_f_g}
   \langle -A_0\mathcal{U}_0^2-A_0V'\mathcal{U}_0+A_0V\mathcal{U}'_0\rangle f_0'' +\langle -\Omega_2^2B_0\mathcal{U}_0^2-\Omega_0^2B_2\mathcal{U}_0^2+A_2(\mathcal{U}_0')^2\rangle f_0 = 0.
\end{equation}
Furthermore, note that when $\Omega_1=0$, $A_2$ and $B_2$ defined in \eqref{eq:def_D2}, regardless of the dispersive model chosen, is: 
\begin{equation}
    \label{rho2_a2_tilde}
    B_2(\Omega_0,0,\Omega_2)=\Omega_2^2\Tilde{B}_2(\Omega_0) \text{ and } A_2(\Omega_0,0,\Omega_2)=\Omega_2^2\Tilde{A}_2(\Omega_0). 
\end{equation}
Therefore, we get the sought-after effective equation for $f_0$ and $\Omega_2$: 
\begin{equation}
    \label{syst_f}
   T f_0'' +\Omega_2^2f_0=0,
\end{equation}
where
\begin{equation}
    \label{def_Ti}
    T = \frac{\langle A_0\mathcal{U}_0^2+A_0V'\mathcal{U}_0-A_0V\mathcal{U}'_0\rangle}{\langle B_0\mathcal{U}_0^2+\Omega_0^2\Tilde{B}_2\mathcal{U}_0^2-\Tilde{A}_2(\mathcal{U}_0')^2\rangle}
    = \frac{\left\langle a(\cdot,\Omega_0)( \mathcal{U}_0^2+ V'\mathcal{U}_0 -V\mathcal{U}'_0)\right\rangle}{\mathcal{S}(\Omega_0,\mathcal{U}_0,\mathcal{U}_0)},
\end{equation}
for any frequency-dependent functions $a$ and $b$, and where for any two functions $f$ and $g$, and any reduced frequency $\Omega_0$, $\mathcal{S}$ is defined by:
\begin{equation}
\mathcal{S}(\Omega_0,f,g) =  \left\langle \left( b(\cdot,\Omega_0)+\frac{\Omega_0}{2}\frac{\partial  b}{\partial \Omega}(\cdot,\Omega_0)\right)f g - \frac{1}{2\Omega_0}\frac{\partial a}{\partial \Omega}(\cdot,\Omega_0) f' g'\right\rangle.
    \label{def_S}
\end{equation}
Therefore, we get the effective string described by \eqref{syst_f} on the long-scale where the complex material properties are now solely concentrated in a single effective parameter $T$ \eqref{def_Ti}. \\
Applying the Bloch-Floquet conditions \eqref{eq:BF_cond} gives the final expression for the quadratic term in the dispersion relation 
\begin{equation}
    \label{Omega_2_final_1D}
    \Omega_2^2= T\tilde{\kappa}^2,
\end{equation}
with $\tilde{\kappa}=\kappa$ near 0 and $\tilde{\kappa}=\pi/\delta-\kappa$ near $\pi/\delta$ so that the dispersion relation is approximated by 
\begin{equation}
    \label{eq:Omega_final_1D}
    \Omega \approx \Omega_0 + \frac{T}{2\Omega_0}(\tilde{\kappa}\delta)^2.
\end{equation}

\subsubsection{Double eigenvalues at the edges of the Brillouin zone}\label{Sec:1D_double}
We next consider the case of multiplicity two for the eigenvalue $\Omega_0$. In that case, $\Omega_1$ is no longer 0, and the zeroth-order wavefield is now written as 
\begin{equation}
    \label{double_1D_u0} u_0(x,\xi)=f_0^{(1)}(x) \mathcal{U}_0^{(1)}(\xi)+f_0^{(2)}(x) \mathcal{U}_0^{(2)}(\xi)
\end{equation}
where $\mathcal{U}_0^{(1)}(\xi)$ and $\mathcal{U}_0^{(2)}(\xi)$ are two independent eigenfunctions associated to $\Omega_0$, while $f_0^{(1)}(x)$ and $f_0^{(2)}(x)$ are the slow modulation functions to be found. \\
Consequently, both eigenfunctions satisfy \eqref{syst_order0_main} and we denote $\eqref{syst_order0_main}^{(i)}$ the equation for the $i$th eigenfunction. Furthermore, because $\Omega_1$ is non-zero, the system satisfied by the first-order field becomes
\begin{equation}
    \label{syst_order1_main_Omega1_non_zero}
A_0\frac{\partial}{\partial\xi}\left(\frac{\partial u_1}{\partial\xi}+2\frac{\partial u_0}{\partial x}\right)+A_1\frac{\partial^2u_0}{\partial\xi^2}+\Omega_0^2(B_0 u_1+B_1 u_0)+\Omega_1^2B_0u_0=0,
\end{equation}
together with periodicity/antiperiodicity for $u_1$, continuity for $u_1$ at $\phi$ and $0$, and continuity for $A_0\left( \frac{\partial u_0}{\partial x}+\frac{\partial u_1}{\partial\xi}\right)+A_1\frac{\partial u_0}{\partial\xi}$ at $\phi$ and $0$.

We introduce $\Tilde{B}_1$ and $\Tilde{A}_1$ so that
\begin{equation}
    \label{rho_a_tilde}
    B_1(\Omega_0,\Omega_1)=\frac{\Omega_1^2}{\Omega_0^2}\Tilde{B}_1(\Omega_0) \text{ and } A_1(\Omega_0,\Omega_1)=\frac{\Omega_1^2}{\Omega_0^2}\Tilde{A}_1(\Omega_0).
\end{equation}
Then considering the equations $\langle\eqref{syst_order0_main}^{(i)}\times u_1 - \eqref{syst_order1_main_Omega1_non_zero}\times \mathcal{U}_0^{(i)}\rangle=0$ for $i=1,2$, 
 we get the effective equation for $\bm{F}=(f_0^{(1)},f_0^{(2)})^\transpose$: 
\begin{equation}
    \label{effective_equation_double_1D}
    \bm{F}'(x) = \frac{\Omega_1^2}{\langle A_0 \mathrm{w}_0\rangle}\bm{\mathcal{N}}\bm{F},
\end{equation}
where $\mathrm{w}_0$ is the Wronskian defined by 
\begin{equation}
    \label{wronskian_1D}
 \mathrm{w}_0(\xi)=  \mathcal{U}_0^{(1)}(\xi){\mathcal{U}_0^{(2)}}'(\xi) -{\mathcal{U}_0^{(1)}}'(\xi)\mathcal{U}_0^{(2)}(\xi),
\end{equation} 
$\bm{\mathcal{N}}$ is the matrix defined by 
\begin{equation}
    \label{matrix_double_1D}
    \bm{\mathcal{N}} = 
    \left(
    \begin{array}{cc}
     \mathcal{S}(\Omega_0,\mathcal{U}_0^{(1)},\mathcal{U}_0^{(2)}) & \mathcal{S}(\Omega_0,\mathcal{U}_0^{(2)},\mathcal{U}_0^{(2)})  \\
     -\mathcal{S}(\Omega_0,\mathcal{U}_0^{(1)},\mathcal{U}_0^{(1)}) & -\mathcal{S}(\Omega_0,\mathcal{U}_0^{(1)},\mathcal{U}_0^{(2)})
    \end{array}
    \right),
\end{equation}
and $\mathcal{S}$ is defined in \eqref{def_S}. 

Regarding the dispersion diagram, using Bloch-Floquet conditions \eqref{eq:BF_cond} we get the two opposite slopes (here the upper and lower notation does not stand for left or right edge of the Brillouin zone but for the upper and lower branch starting from $\Omega_0$)
\begin{equation}
    \label{Omega1_double_1D}
    \Omega_1^2= \pm T_D \tilde{\kappa}
\end{equation}
so that
\begin{equation}
    \label{Omega_double_1D}
    \Omega \approx \Omega_0\pm \frac{T_D}{2\Omega_0}\tilde{\kappa}\delta
\end{equation}
with $\tilde{\kappa}=\kappa$ near 0 and $\tilde{\kappa}=\pi/\delta-\kappa$ near $\pi/\delta$, and $T_D$ defined by 
\begin{align}
    \label{Td_1D}
    T_D = \frac{\langle a(\cdot,\Omega_0) \mathrm{w}_0\rangle}{\left(
    \mathcal{S}(\Omega_0,\mathcal{U}_0^{(1)},\mathcal{U}_0^{(1)}) \mathcal{S}(\Omega_0,\mathcal{U}_0^{(2)},\mathcal{U}_0^{(2)}) 
    -\mathcal{S}(\Omega_0,\mathcal{U}_0^{(1)},\mathcal{U}_0^{(2)}) ^2
    \right)^{1/2}}.
\end{align}
\subsubsection{Nearby eigenvalues at the edges of the Brillouin zone \label{sec:nearby1Dedge}}
Let us assume now that we have two simple eigenvalues close to each other following \cite{Assier2020,Guzina2019}. The two nearby simple eigenvalues are denoted by $\Omega_0^{(1)}$ and $\Omega_0^{(2)}$. Their proximity is quantified by writing
\begin{equation}
    \label{introduce_distance}
   (\Omega_0^{(2)})^2-(\Omega_0^{(1)})^2= \alpha\delta,
\end{equation}
for some constant $\alpha>0$.
We denote by $\mathcal{U}_0^{(1)}$ and $\mathcal{U}_0^{(2)}$ the eigenfunctions associated to $\Omega_0^{(1)}$ and $\Omega_0^{(2)}$, respectively. \\
The ansatz is considered around the eigenvalue $\Omega_0^{(1)}$:
\begin{eqnarray}
  u_{\delta} (x) = \sum_{j \geqslant 0} \delta^j u_j (x, \xi) & \text{ \ and \ } &
  \Omega^2 = (\Omega_0^{(1)})^2+\sum_{\ell \geqslant 1} \delta^{\ell} \Omega_{\ell}^2, 
\label{eq:asymptoticrepfield_nearby}
\end{eqnarray}

By similarity with the double eigenvalue case, and to take into account the coupling between both eigenvalues we look for the zeroth-order wavefield as 
\begin{equation}
    \label{decompo_zero_nearby}
    u_0(x,\xi)= f_0^{(1)}(x) \mathcal{U}_0^{(1)}(\xi)+f_0^{(2)}(x) \mathcal{U}_0^{(2)}(\xi).
\end{equation}
We will use the notation $D_j^{(i)}$, for $D=A,\,B$ and 
$i=1,\,2$, and $j=0,\,1$ so that we get from Taylor expansions
\begin{equation}
    \label{zero_eigenfunction_nearby}
A_0^{(i)}\frac{\partial^2\mathcal{U}_0^{(i)}}{\partial\xi^2}+(\Omega_0^{(i)})^2B_0^{(i)}\mathcal{U}_0^{(i)} = 0
\end{equation}
and
\begin{equation}
    \label{D_0^2}
    D_0^{(2)} = D_0^{(1)}+\frac{\alpha\delta}{2\Omega_0^{(1)}}\frac{\partial d}{\partial\Omega}(\Omega_0^{(1)})+\mathcal{O}(\delta^2)
\end{equation}
where $d=a,\,b$ for $D=A,\,B$, respectively.
This leads to  
\begin{equation}
    \label{eq_u0_nearby_with_remainder}
  \begin{multlined}
    A_0^{(1)}\frac{\partial^2u_0}{\partial\xi^2}+(\Omega_0^{(1)})^2B_0^{(1)}u_0 = \delta \left[\frac{\alpha}{2A_0^{(1)}} \Omega_0^{(1)}B_0^{(1)}\frac{\partial a}{\partial\Omega}(\xi,\Omega_0^{(1)}) \right.\\ \left. -\alpha B_0^{(1)}-\frac{\alpha}{2}\Omega_0^{(1)}\frac{\partial b}{\partial\Omega}(\xi,\Omega_0^{(1)})\right]u_0^{(2)}.
    \end{multlined}
\end{equation}
Consequently, \eqref{syst_order0_main} is satisfied by the zeroth-order wavefield up to right-hand side residual term of \eqref{eq_u0_nearby_with_remainder} that modifies the equation for the first order field that now becomes: 
\begin{equation}
    \label{syst_order1_nearby}
    \begin{aligned}
&A_0^{(1)}\frac{\partial}{\partial\xi}\left(\frac{\partial u_1}{\partial\xi}+2\frac{\partial u_0}{\partial x}\right)+A_1^{(1)}\frac{\partial^2u_0}{\partial\xi^2}+(\Omega_0^{(1)})^2(B_0^{(1)} u_1+B_1^{(1)} u_0)+\Omega_1^2B_0^{(1)}u_0\\
&+\alpha u_0^{(2)}\left[\frac{1}{2A_0^{(1)}} \Omega_0^{(1)}B_0^{(1)}\frac{\partial a}{\partial\Omega}(\xi,\Omega_0^{(1)})- B_0^{(1)}-\frac{1}{2}\Omega_0^{(1)}\frac{\partial b}{\partial\Omega}(\xi,\Omega_0^{(1)}) \right]=0,
\end{aligned}
\end{equation}
together with periodicity/antiperiodicity for $u_1$, continuity for $u_1$ at $\phi$ and $0$, and continuity for $A_0^{(1)}\left( \frac{\partial u_0}{\partial x}+\frac{\partial u_1}{\partial\xi}\right)+A_1^{(1)}\frac{\partial u_0}{\partial\xi}$ at $\phi$ and $0$.
As in the double eigenvalue case, considering  $\langle \eqref{zero_eigenfunction_nearby}^{(i)}\times u_1 - \eqref{syst_order1_nearby}\times \mathcal{U}_0^{(i)}\rangle$ for $i=1,2$  allows to obtain the effective equation for $\bm{F}=(f_0^{(1)},f_0^{(2)})^\mathrm{T}$: 
\begin{equation}
    \label{effective_equation_nearby_1D}
    \bm{F}'(x) = \frac{\Omega_1^2}{\langle A_0^{(1)} \mathrm{w}_0\rangle}\bm{\mathcal{N}}_\alpha\bm{F}(x)
\end{equation}
with $\mathrm{w}_0$ still given by \eqref{wronskian_1D}. The matrix
$\bm{\mathcal{N}}_\alpha$ is defined by
\begin{equation}
    \label{matrix_double_1D_nearby}
    \bm{\mathcal{N}}_\alpha =     \left(
    \begin{array}{cc}
     \mathcal{S}(\Omega_0^{(1)},\mathcal{U}_0^{(1)},\mathcal{U}_0^{(2)})  & \mathcal{S}(\Omega_0^{(1)},\mathcal{U}_0^{(2)},\mathcal{U}_0^{(2)}) \\
     -\mathcal{S}(\Omega_0^{(1)},\mathcal{U}_0^{(1)},\mathcal{U}_0^{(1)}) & -\mathcal{S}(\Omega_0^{(1)},\mathcal{U}_0^{(1)},\mathcal{U}_0^{(2)}) 
    \end{array}
    \right)+\frac{\alpha}{\Omega_1^2}\left(
    \begin{array}{cc}
     0 & -\mathcal{G}_2 \\
     0 &  \mathcal{G}_1
    \end{array}
    \right)
\end{equation}
with 
\begin{equation}
    \label{def_Bi_Ci_nearby}
    \begin{aligned}
\mathcal{G}_i &= \left\langle (\tilde{B}_1^{(1)}+B_0^{(1)}) \mathcal{U}_0^{(i)}\mathcal{U}_0^{(2)}  -\frac{B_0^{(1)}}{A_0^{(1)}}\tilde{A}_1^{(1)}\mathcal{U}_0^{(i)}\mathcal{U}_0^{(2)}\right\rangle \\
& =\left\langle b(\cdot,\Omega_0^{(1)})\left( 1 + \frac{\Omega_0^{(1)}}{2} \frac{\partial }{\partial\Omega}[\log (b/a)](\cdot,\Omega_0^{(1)}) \right)\mathcal{U}_0^{(i)}\mathcal{U}_0^{(2)}\right\rangle.
\end{aligned}
\end{equation}
The dispersion relation is then obtained by solving
\begin{equation}
    \label{det_disp_nearby}
    \mathrm{det}\left( \frac{\Omega_1^2}{\langle A_0^{(1)} \mathrm{w}_0\rangle}\bm{\mathcal{N}}_\alpha-\mathrm{i}\kappa\mathbb{I}_2\right) =0.
\end{equation}
One notes that $\bm{\mathcal{N}}_\alpha=\bm{\mathcal{N}}$ when $\alpha=0$ and we then recover the double case.
\subsubsection{Simple eigenvalues inside the Brillouin zone (no damping)}\label{Sec:1D_inside}
In this section, we get a linear approximation for an arbitrary point inside (strictly) the Brillouin zone. However, this is possible only if the physical parameters are real. Consequently, we consider all the damping terms $\gamma_{p,1}$ and $\gamma_{p,2}$ equal to 0, in the framework of this subsection only, so that the coefficients $A_i$ and $B_i$ ($i=1,2$) are real. \\
Let us pick a point $(\kappa^\star,\Omega_0)$ with $\kappa^\star \notin\left\lbrace0,\pi/\delta\right\rbrace$ and $\Omega_0$ solution of the eigenvalue problem satisfied by the zeroth-order field:
\begin{equation}
    \label{inside_1D_zero}
A_0\frac{\partial^2u_0}{\partial\xi^2}+\Omega_0^2B_0u_0 = 0 \text{ in } (0,\delta)\times(0,1),
\end{equation}
together with $u_0(x,\xi+1)=\mathrm{e}^{\mathrm{i}\kappa^\star\delta}u_0(x,\xi)$ in $(0,\delta)\times(0,1) $, continuity for $u_0$ at $\xi=\phi$ and $\xi=0$, and continuity for $A_0 \frac{\partial u_0}{\partial \xi}$ at $\xi=\phi$ and $\xi=0$. \\
For the first order, we get: 
\begin{equation}
    \label{inside_1D_first}
           A_0\frac{\partial}{\partial\xi}\left(\frac{\partial u_1}{\partial\xi}+2\frac{\partial u_0}{\partial x}\right)+A_1\frac{\partial^2u_0}{\partial\xi^2}+\Omega_0^2(B_0 u_1+B_1 u_0)+\Omega_1^2B_0u_0=0 \text{ in } (0,\delta)\times(0,1),
\end{equation}
together with $u_1(x,\xi+1)=\mathrm{e}^{\mathrm{i}\kappa^\star\delta}u_1(x,\xi)$ in $(0,\delta)\times(0,1) $, continuity for $u_1$ at $\xi=\phi$ and $\xi=0$, and continuity for $A_0\left( \frac{\partial u_0}{\partial x}+\frac{\partial u_1}{\partial\xi}\right)+A_1\frac{\partial u_0}{\partial\xi}$ at $\xi=\phi$ and $\xi=0$. \\
We consider $I=\langle \bar{u}_1\times\eqref{inside_1D_zero}-\bar{\eqref{inside_1D_first}}\times u_0\rangle=0$. We still have 
\begin{equation}
    \left\langle \frac{\partial}{\partial\xi}\left[ A_0\left( \frac{\partial \bar{u}_0}{\partial x}+\frac{\partial \bar{u}_1}{\partial\xi}\right)u_0+A_1\frac{\partial \bar{u}_0}{\partial\xi}u_0 -A_0\frac{\partial u_0}{\partial\xi}\bar{u}_1\right] \right\rangle = 0
\end{equation}
because the quantities above are continuous and 1-periodic. Therefore,  $I=0$ reduces to 
\begin{equation}
   \left \langle A_0 \left( \frac{\partial^2\bar{u}_0}{\partial\xi\partial x}u_0-\frac{\partial\bar{u}_0}{\partial x}\frac{\partial u_0}{\partial\xi}\right)-A_1\left|\frac{\partial u_0}{\partial\xi}\right|^2+\Omega_0^2 
  B_1 |u_0|^2+\Omega_1^2B_0|u_0|^2\right\rangle = 0.
\end{equation}
Dividing through by $f_0$ and using \eqref{inside_1D_zero}, we end up with a first-order ODE for $f_0$:
\begin{equation}
    \label{effective_eq_inside_1D}
    T_i f_0'(x)-\mathrm{i}\Omega_1^2f_0(x)=0
\end{equation}
with $T_i$ defined by 
\begin{equation}
    \label{T_inside_no_damping}
    T_i 
     = \frac{2\langle a(\cdot,\Omega_0) \Im(\mathcal{U}_0'\bar{\mathcal{U}}_0) \rangle}{\mathcal{S}(\Omega_0,\mathcal{U}_0,\bar{\mathcal{U}}_0)},
\end{equation}
where we remind that $S$ is defined in \eqref{def_S}.
Regarding the dispersion relation, applying the Bloch-Floquet conditions \eqref{eq:BF_cond} gives
\begin{equation}
    \label{DD_1D_inside}
    \Omega_1^2=T_i(\kappa-\kappa^\star).
\end{equation}

\subsubsection{The low-frequency case}
  \label{eq:asymptoticrepfield}
We also obtain the classical low-frequency homogenization by considering $\Omega_0=0$ in \eqref{syst_order0_main}, which leads to the fact that $\mathcal{U}_0$ is uniform, and we, without loss of generality, choose $\mathcal{U}_0=1$. Then, we write $u_1(x,\xi)=f_0'(x)\mathcal{U}_1(\xi)$ where $u_1$ satisfies \eqref{syst_order1_main} with $\Omega_0=0$ and $u_0=f_0$. Therefore $\langle A_0(1+\mathcal{U}_1')\rangle=\langle 1/A_0\rangle^{-1}$. Integration on a unit cell of \eqref{syst_order2_main} for $\Omega_0=0$ then leads to the usual homogenized equation
\begin{equation}
    \label{eq:LF_limit}
    \langle 1/A_0\rangle^{-1} f''_0(x)+\Omega_2^2\langle B_0\rangle f_0(x) = 0
\end{equation}
and dispersion relation
\begin{equation}
    \label{eq:LF_limit_DD}
    \Omega_2^2=\langle 1/A_0\rangle^{-1}\langle B_0\rangle^{-1}\kappa^2.
\end{equation}

\subsection{Numerical investigation}
\label{sec:numerical}

We now use two different methods to compute the whole dispersion diagrams: we either track the zeros of the dispersion function \eqref{eq:Disp} in the complex plane along a branch, or we use the finite element method (FEM) to directly solve for \eqref{eq:Raph-BF-ODEs}. The details are given in Appendix \ref{Sec:App2} for the latter. Hereafter, the dispersion diagrams computed either by zero tracking or by FEM from the exact dispersion function will be denoted as the \textit{exact dispersion diagrams} in contrast to the asymptotic approximations obtained by HFH with which they will be compared. 

\subsubsection{Dielectric and metallic layers (Drude with damping)}
\begin{figure}[h!]
\subfloat[Phase portrait in complex plane for the dispersion (i.e. 
 determinant) function \eqref{eq:Disp} taken at $\kappa=0$]{\includegraphics[trim = 0mm 0mm 0mm 0mm, clip, width = 0.45\linewidth]{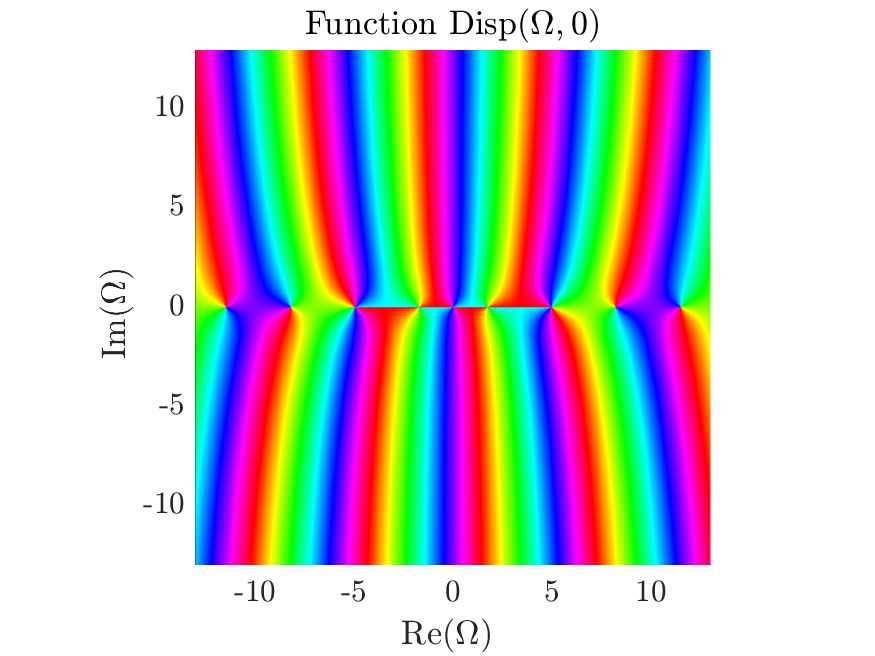}\label{fig:complex_map_silver_a}}\hspace{5pt}
\subfloat[Dispersion diagram in the complex plane \\(here $\kappa\in(0\pi/\delta)$)]{\hspace{-10pt}\includegraphics[trim = 0mm 0mm 0mm 0mm, clip, width = 0.55\linewidth]{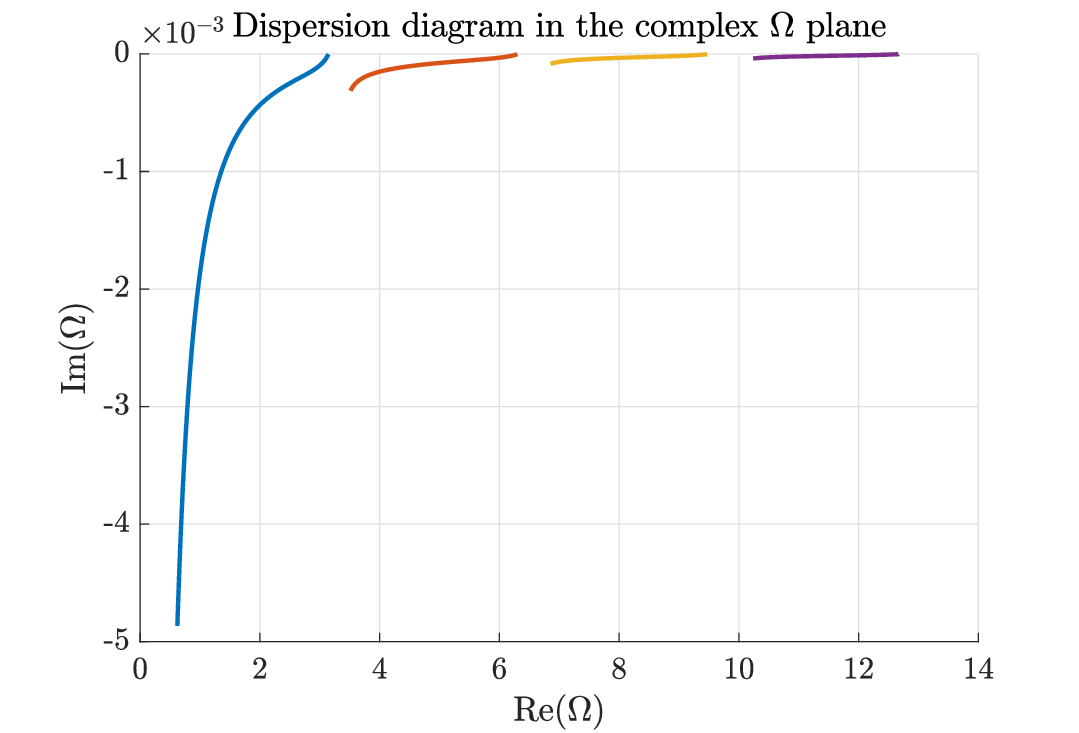}\label{fig:complex_map_silver_b}}
\caption{Representation of the dispersion function and its zeros.}
\label{fig:complex_map_silver}
\end{figure}

Motivated by the configuration of \cite{ZapataRodriguez2014}, we consider wave propagation through alternated layers of silver (Ag) and Titanium dioxide (TiO$_2$). Here only the permittivity is frequency-dependent, following a Drude law in the metal layer made of silver (Ag). More precisely, we have $a=1$ in both materials, and in  silver  
$$\mathbb{b}(\Omega)=\frac{b_\infty}{b_0}\left(1-\frac{\Omega_{1,2}^2}{\Omega(\Omega+\mathrm{i}\gamma_{1,2})}\right)$$
with $b_0=6.2$, $ b_\infty=1$, and one resonance for $ b$ with $\Omega_{1,2}=5.01$, and $\gamma_{1,2}=0.01$ in \eqref{rho_adim}. The filling ratio of the dielectric layer is $\phi=10/11$ for a periodicity of $h=110$ nm. \\
Since there is damping, the frequency solutions of the dispersion relation are complex and to visualize the dispersion function and its zeros without having to consider the real and imaginary parts separately, we plot at a given frequency its phase portrait in the complex plane, see Figure \ref{fig:complex_map_silver_a} for $\kappa=0$. Alternatively, we track the zeros of the function along a branch of the dispersion diagram in the complex plane, see Figure \ref{fig:complex_map_silver_b}.

\begin{figure}[h!]
\subfloat[Real part]{\includegraphics[trim = 0mm 0mm 0mm 0mm, clip, width = 0.45\linewidth]{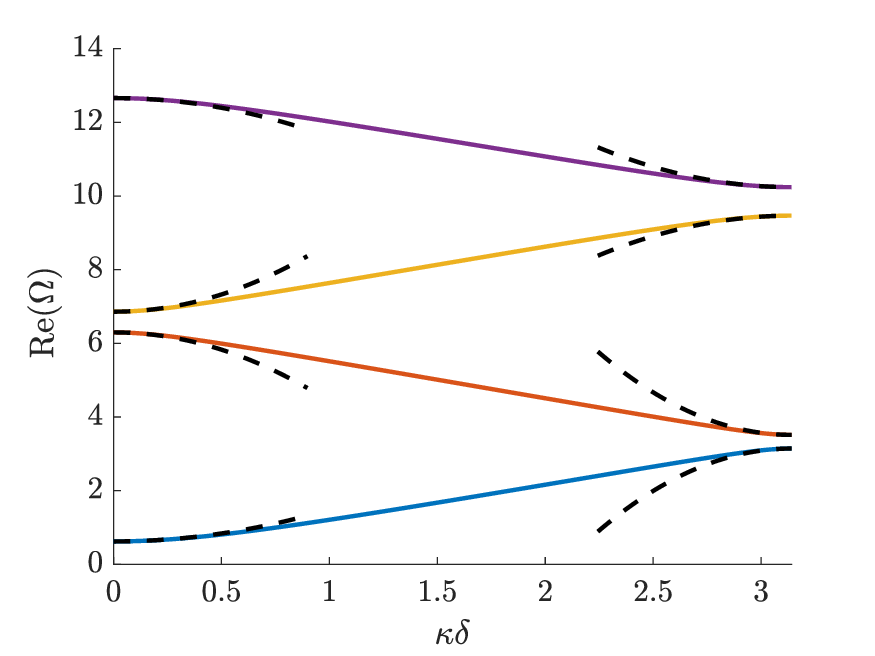}}
\hspace{0.5cm}
\subfloat[Imaginary part]{\includegraphics[trim = 0mm 0mm 0mm 0mm, clip, width = 0.45\linewidth]{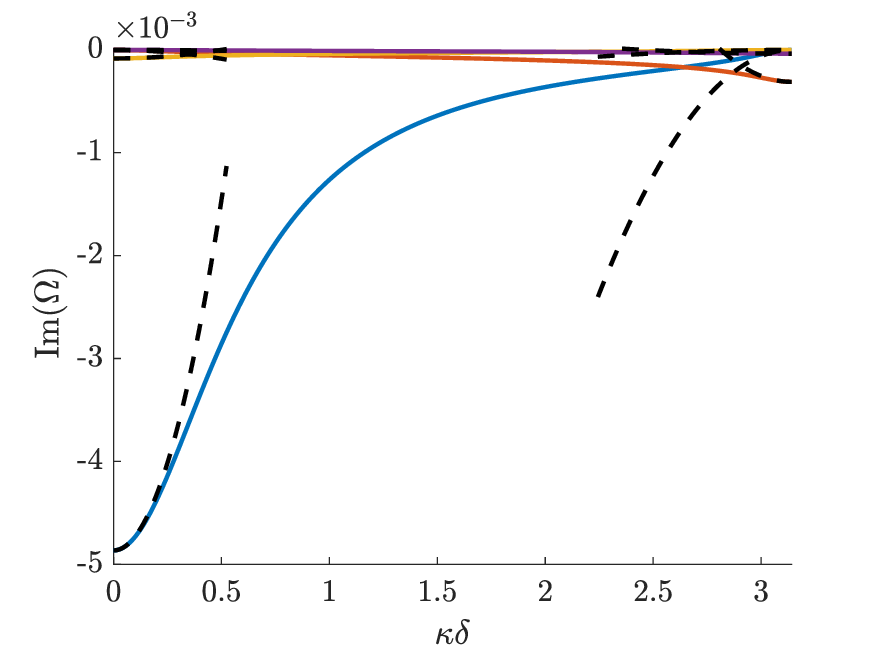}}
\caption{Superposition of the dispersion diagram from zero tracking (solid lines) and the asymptotic approximations obtained by HFH in the simple eigenvalue case \eqref{eq:Omega_final_1D} around $\tilde{\kappa}=0$ (dashed lines).}
\label{fig:comp_silver}
\end{figure}

\begin{figure}[h!]
\subfloat[Near $\kappa\delta=0$]{\includegraphics[trim = 0mm 0mm 0mm 0mm, clip, width = 0.4\linewidth]{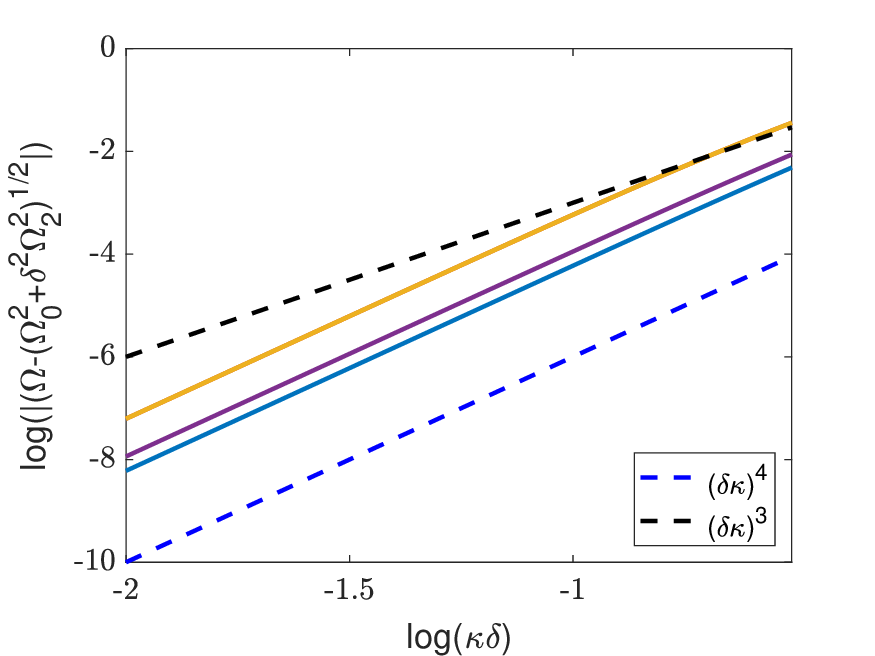}}
\hspace{0.5cm}
\subfloat[Near $\kappa\delta=\pi$]{\includegraphics[trim = 0mm 0mm 0mm 0mm, clip, width = 0.45\linewidth]{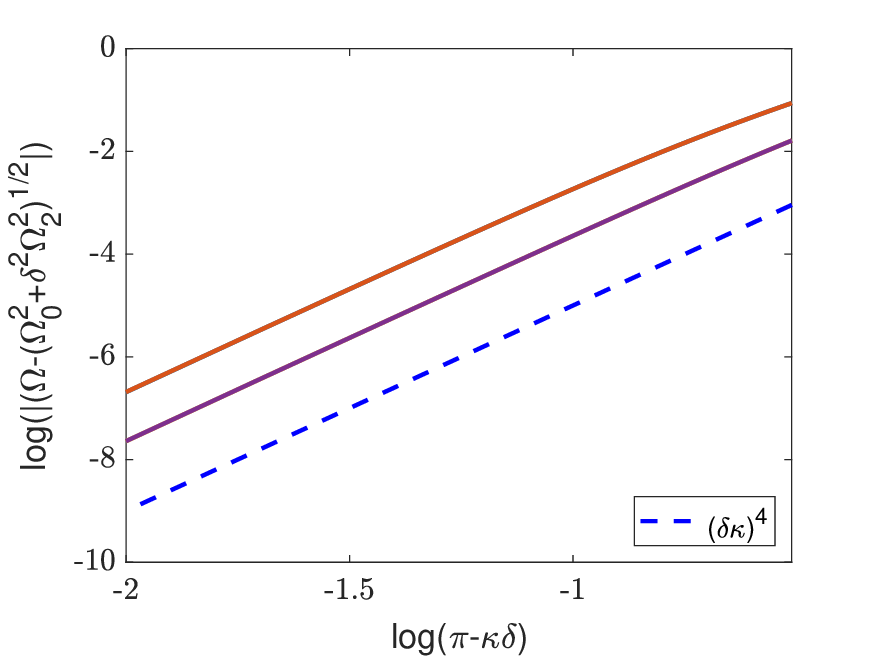}} 
\caption{Absolute error between the exact dispersion diagram and the HFH approximation in the simple eigenvalue case \eqref{eq:Omega_final_1D} in a log-log scale (solid lines). Dashed lines are reference orders of convergence. }
\label{fig:conv_silver}
\end{figure}

\paragraph{Simple eigenvalue approximations at the edges}
Firstly, we use the asymptotic approximations obtained for simple eigenvalues near the edges of the Brillouin zone, see Section \ref{Sec:1D_simple}. The resulting asymptotic approximations of the dispersion diagrams for both the real part and the imaginary part are displayed in  Figure \ref{fig:comp_silver}, where we used the zero tracking method to compute the diagrams for the exact dispersion relation. The absolute errors for each of the branches and for both $\kappa=0$ and $\kappa=\pi/\delta$ are then shown in Figure \ref{fig:conv_silver}, where we recover that the quadratic term $\Omega_2$ is well taken into account asymptotically and that in fact the next term is zero so that the error is $\mathcal{O}((\tilde{\kappa}\delta)^4)$.

\begin{figure}[h!]
\subfloat[Real part for the first and second mode around $\kappa\delta = \pi$.\label{fig:nearby_Zapata_a}]{\includegraphics[trim = 0mm 0mm 0mm 0mm, clip, width = 0.45\linewidth]{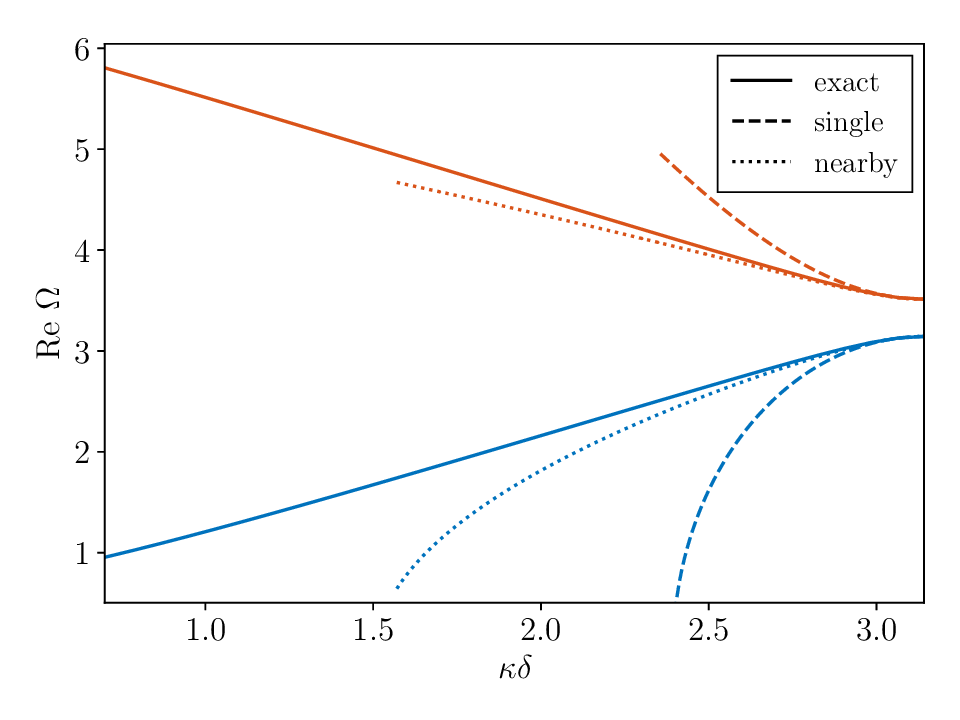}}
\hspace{0.5cm} 
\subfloat[Imaginary part for the first and second mode around $\kappa\delta = \pi$.\label{fig:nearby_Zapata_b}]{\includegraphics[trim = 0mm 0mm 0mm 0mm, clip, width = 0.45\linewidth]{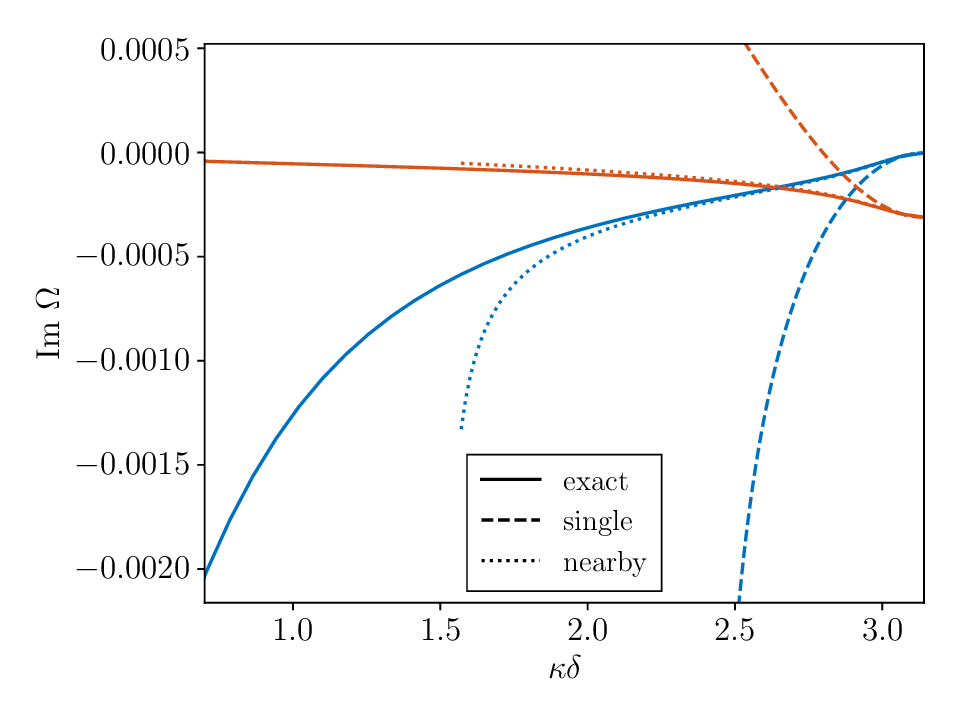}}  \\ 
\subfloat[Real part for the second and third mode around $\kappa\delta = 0$.\label{fig:nearby_Zapata_c}]{\includegraphics[trim = 0mm 0mm 0mm 0mm, clip, width = 0.45\linewidth]{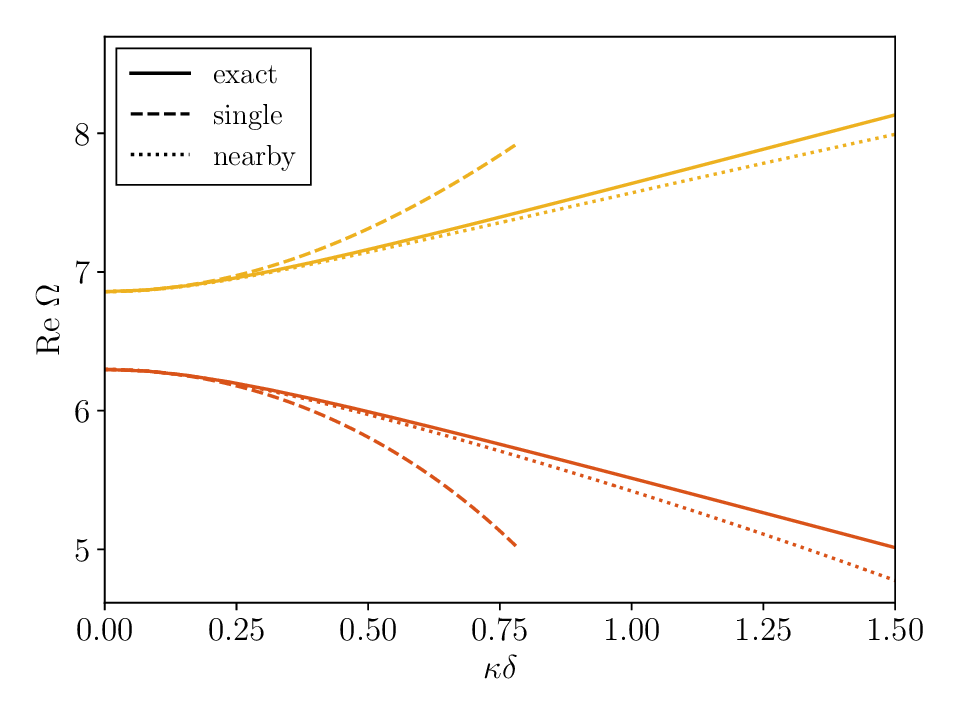}}
\hspace{0.5cm}
\subfloat[Imaginary part for the second and third mode around $\kappa\delta = 0$.\label{fig:nearby_Zapata_d}]{\includegraphics[trim = 0mm 0mm 0mm 0mm, clip, width = 0.45\linewidth]{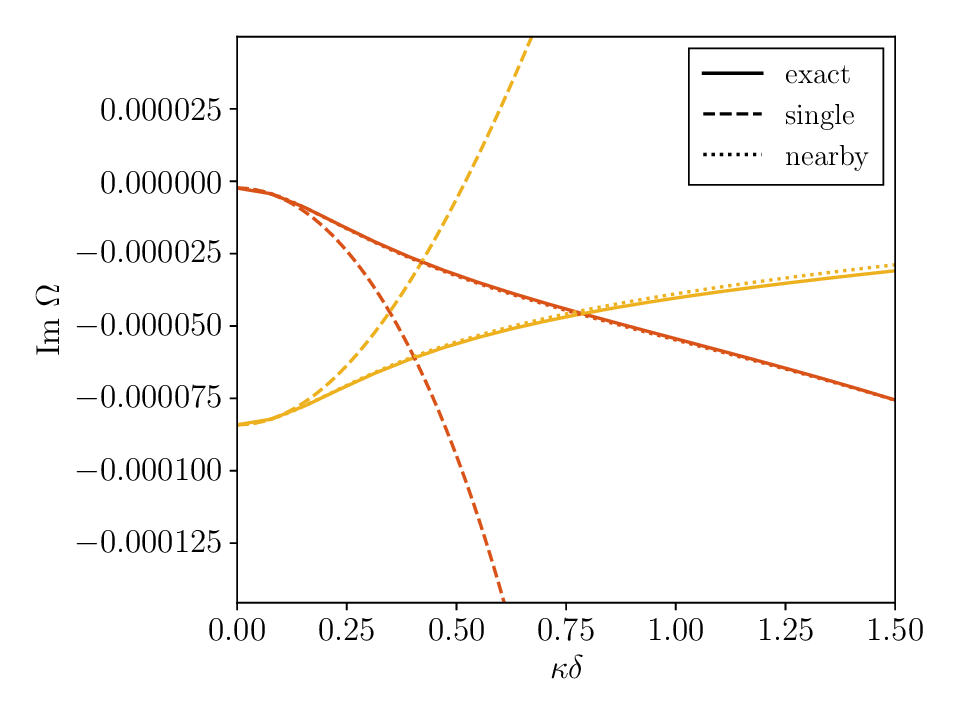}} 
\caption{Comparison of the exact dispersion relation (solid lines) and the effective one obtained by HFH in the simple eigenvalue case \eqref{eq:Omega_final_1D} (dashed lines) and with nearby approximations \eqref{det_disp_nearby} (dotted lines).}
\label{fig:nearby_Zapata}
\end{figure}

\begin{figure}[h!]
\includegraphics[trim = 0mm 0mm 0mm 0mm, clip, width = 0.9\linewidth]{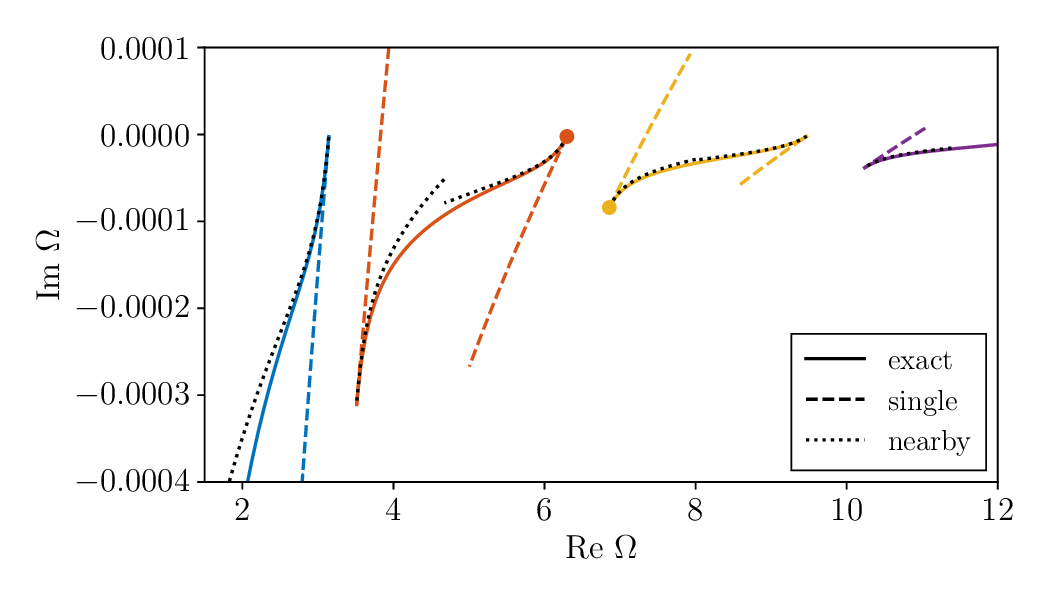} 
\caption{Comparison in the complex plane of the exact dispersion relation (solid lines) and the effective one obtained by HFH for the single eigenvalue case \eqref{eq:Omega_final_1D} (dashed lines) and with nearby approximations \eqref{det_disp_nearby} (dotted lines). The circle markers correspond to $\kappa\delta=0$.}
\label{fig:nearby_Zapata_complex}
\end{figure}

\paragraph{Nearby approximations at the edges}
Given these numerical solutions, we now compare with the asymptotics and 
 use the nearby approximations developed in Section \ref{sec:nearby1Dedge} for single eigenvalues near the edges of the Brillouin zone. Solutions from the FEM method are shown in Figure \ref{fig:nearby_Zapata} for real and imaginary parts for four modes, and in Figure \ref{fig:nearby_Zapata_complex} for the dispersion diagram in the complex plane. It is clearly
seen that the agreement with the dispersion diagram is much longer lived than that of the simple
eigenvalue approximation for both real and imaginary parts. The nearby approximation leads notably to a better fit of the imaginary parts, which are quite small due to the fact that the damping coefficients $\gamma_i$ are also small in practice. 

\subsubsection{Stack of positive and negative index materials (Lorentz with no damping)}
\begin{figure}
    \centering
      \subfloat[Map in $(\kappa,\Omega)$ space of the logarithm]{  \includegraphics[width=0.45\linewidth]{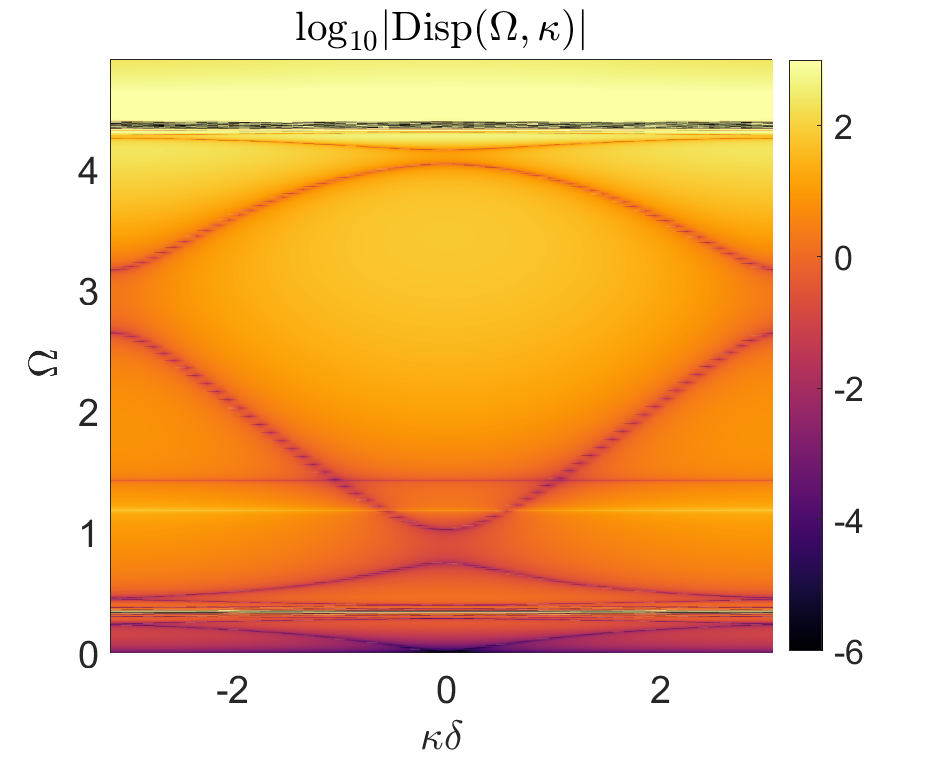}\label{fig:Sheng_det_a}}
  \subfloat[Phase portrait around an accumulation point at $\kappa=0$]{  \includegraphics[width=0.5\linewidth]{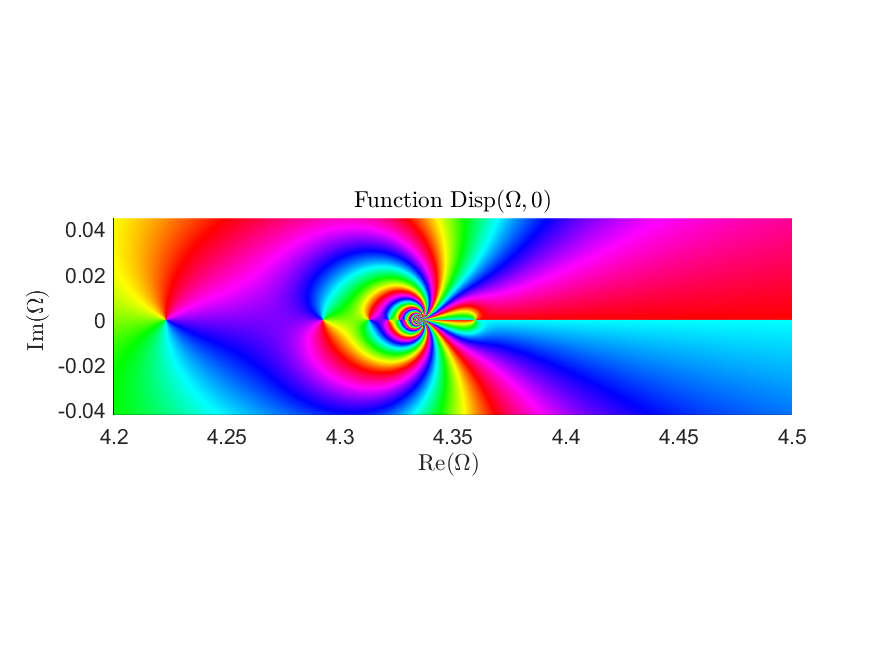}\label{fig:Sheng_det_b}}    
    \caption{Dispersion function. (Left) Map of the logarithm of the dispersion function in the wavenumber-frequency space. The zeros are the solutions of the dispersion relation, therefore represented by the dark lines. (Right) Phase portrait in the complex plan zoomed around the accumulation point which occurs at $\mathcal{R}^+_{1,1}=\Omega_{D,1,1}=4.3354$, as predicted in Section \ref{Sec:BlochFloquet}}
    \label{fig:Sheng_det}
\end{figure}

In a second more challenging example, we now reproduce the results of Li et al. \cite{Li2003}, see Figure 2 of the latter and compare with the asymptotic results. This consists of a 1D system of periodicity $h=18$ mm with alternate layers of air (12 mm thick) and of an effective material which is dispersive (6.0 mm thick). Both parameters in the dispersive medium follow a Lorentz law without damping in the effective layers, see \eqref{a_adim} and \eqref{rho_adim}. More precisely, we set $\phi=2/3$, $a_{\infty} =a_0= b_{\infty} = b_0 =1$, $\Omega_{1,1}=1.131$, $\Omega_{D,1,1}=0.34$, $\Omega_{1,2}=1.885$, $\Omega_{D,1,2}=0.3393$, $\Omega_{2,2}=3.7699$ and $\Omega_{D,2,2}=4.3354$ resulting in 
\begin{equation}
    \label{effective_parameters_Li}
    \begin{aligned}
        & \mathbb{a}(\Omega) = \left(1+\frac{\Omega_{1,1}^2}{\Omega^2-\Omega_{D,1,1}^2} \right)^{-1} \\
        &  \mathbb{b}(\Omega) = 1+\frac{\Omega_{1,2}^2}{\Omega^2-\Omega_{D,1,2}^2}+\frac{\Omega_{2,2}^2}{\Omega^2-\Omega_{D,2,2}^2}.
    \end{aligned}
\end{equation}
We first compute the dispersion relation using Bloch-Floquet analysis (see Section \ref{Sec:BlochFloquet}). The logarithm of the dispersion function \eqref{eq:Disp}, is plotted in Figure \ref{fig:Sheng_det_a}, the dispersion curve therefore corresponds to the dark lines in the map.  The main features of Figure 2 in \cite{Li2003} are recovered, together with the appearance of accumulation points, see Figure \ref{fig:Sheng_det_b} for phase portraits zoomed-in around one of these points. However, we will consider the same range of frequencies as in \cite{Li2003}, for which we are away from any of these points and able to propose high-frequency homogenized approximations. 

\paragraph{Simple eigenvalue approximations at the edges}

\begin{figure}[h!]
\subfloat[Superposition of the exact dispersion diagram (plain lines) and the asymptotic approximations obtained by HFH around the edges (dotted lines)]{\includegraphics[trim = 0mm 0mm 0mm 0mm, clip, width = 0.45\linewidth]{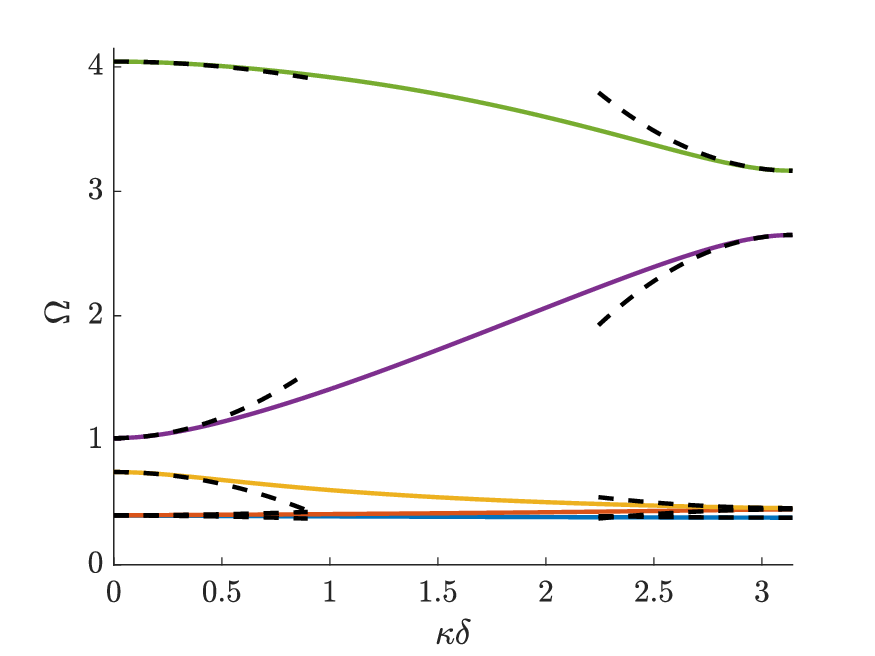}\label{fig:conv_Sheng_a}}
\hspace{0.5cm}
\subfloat[Absolute error near $\kappa\delta=0$ between both dispersion relations in a log-log scale (plain lines). Dotted lines are reference orders of convergence.]{\includegraphics[trim = 0mm 0mm 0mm 0mm, clip, width = 0.45\linewidth]{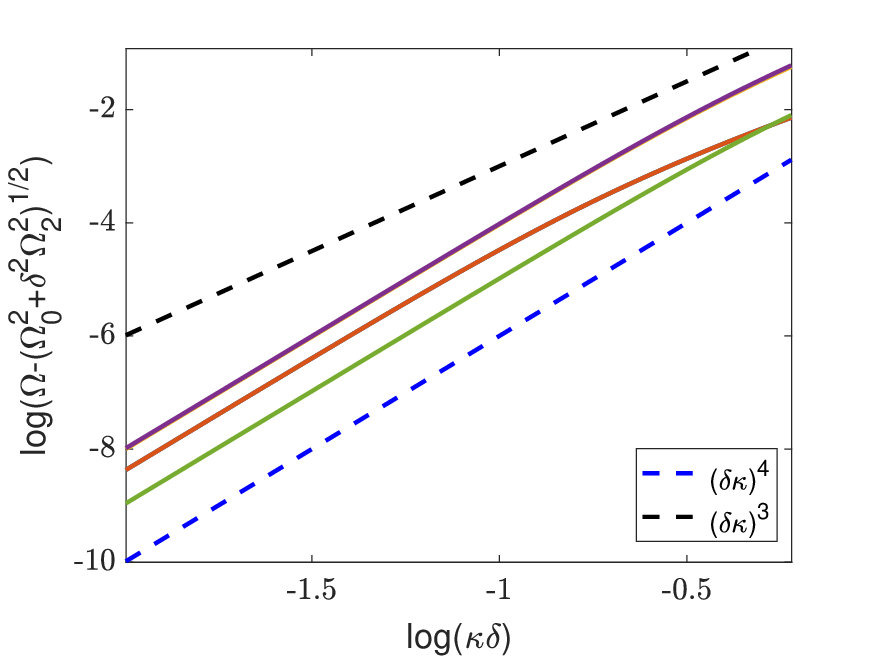}\label{fig:conv_Sheng_b}} 
\caption{Comparison of the exact dispersion relation and the effective one obtained by HFH for the single eigenvalue case \eqref{eq:Omega_final_1D}.}
\label{fig:conv_Sheng}
\end{figure}

We then recover this band diagram near the edges using the quadratic approximations obtained with HFH in Section \ref{Sec:1D_simple}. The comparison is given in Figure \ref{fig:conv_Sheng_a} where the numerical curves are obtained using zero tracking; the branches near the edges are well approximated. For a quantitative validation, we plot the difference between the exact dispersion relation and the  quadratic approximation on a log-log scale. It validates the approximation of the quadratic term and again underlines that there is no third-order term because we get an error of order $\mathcal{O}((\delta\tilde{\kappa})^4)$, see Figure \ref{fig:conv_Sheng_b} for the case $\kappa=0$. \\

\begin{figure}[h!]
\subfloat[Periodic source located at $\xi=0.8$ with frequency close to the fourth eigenfrequency in Fig.(\ref{fig:conv_Sheng}) for $\kappa\delta=0$ ($\Omega=1.01\Omega_0^{(4)}=1.027$). The effective coefficient is $T = 1.336$.\label{fig:reconstr_1D_periodic}]{\includegraphics[trim = 0mm 0mm 0mm 0mm, clip, width = 0.45\linewidth]{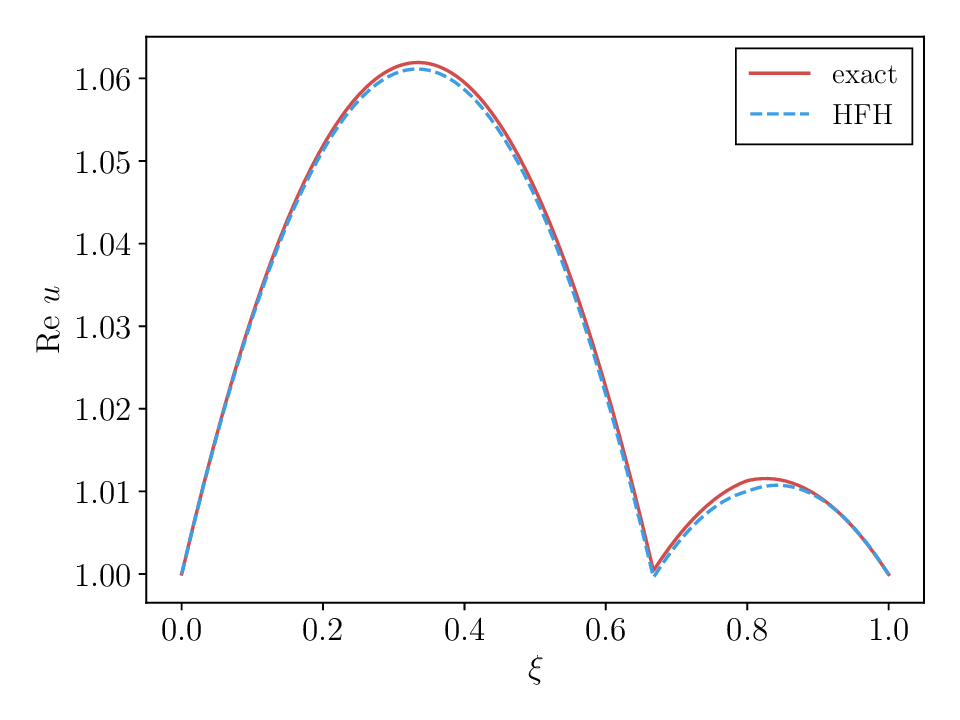}}
\hspace{0.5cm}
\subfloat[Finite stack of 20 periods with a source located at $\xi=10+\phi/2$ with frequency close to the third eigenfrequency in Fig.(\ref{fig:conv_Sheng}) for $\kappa\delta=0$ ($\Omega=1.01\Omega_0^{(3)}=0.752$). The dashed line shows the envelope $f_0$ obtained by HFH. The effective coefficient is $T = -0.573$.\label{fig:reconstr_1D_finite}]{\includegraphics[trim = 0mm 0mm 0mm 0mm, clip, width = 0.45\linewidth]{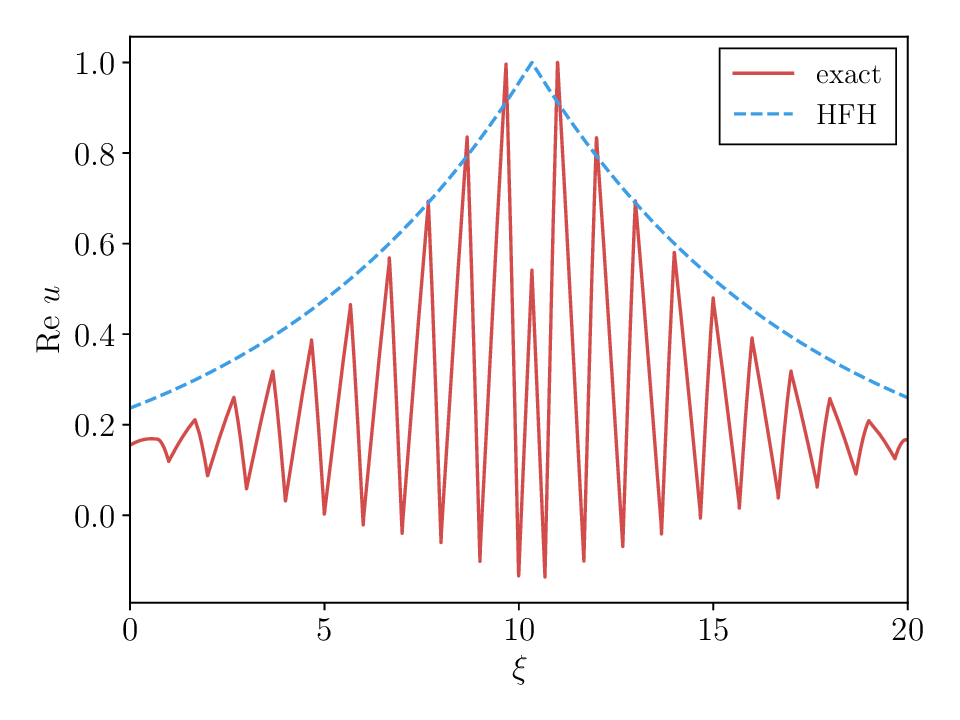}} 
\caption{Comparison of the wavefields for a point source forcing. HFH approximation is the one of the simple eigenvalue case \eqref{syst_f}.}
\label{fig:reconstr_1D}
\end{figure}

Obtaining accurate asymptotics for the dispersion curves is a useful application of the theory, by validating it and by encapsulating the physics into a coefficient $T$ that allows us to tune or design features. An equally important application of the theory is to model forcing, that is, to apply a source in a structured medium and then use the effective equations to model the response; we now proceed to demonstrate the efficiency of that approach. 
We introduce a source term and choose a frequency of excitation close to an eigenfrequency at $\kappa=0$, and then compare the wavefields for the microstructured medium using both numerical simulation and the high-frequency approximations in Figure \ref{fig:reconstr_1D}. We first use a point source spatially located at $\xi =0.8$ and repeated periodically, with frequency $\Omega=1.01\Omega_0^{(4)}=1.027$, close to the fourth eigenfrequency studied in Fig.~(\ref{fig:conv_Sheng_a}); numerically this is modelled by finite elements studying one unit cell and applying periodic boundary conditions. As a comparison we solve the effective equation (\ref{syst_f}) obtained by HFH to get the envelope function $f_0$ and then recover the first order field using Eq. (\ref{decompo_u0_single}). As displayed on Fig.~(\ref{fig:reconstr_1D_periodic}), an excellent agreement is obtained with the simulations for the microstructured medium  (solid lines) and the homogenized one (dashed lines). Next, we consider a finite stack consisting of 20 periods of the microstructured medium and compare it with the effective medium. The point source is located in the center at $\xi=10+\phi/2$ with frequency $\Omega=1.01\Omega_0^{(3)}=0.752$, and numerically we use Perfectly Matched Layers \cite{berenger1994perfectly} on either side to truncate the simulation domain and damp propagating waves to avoid reflections at the computational domain boundary. 
The dashed lines on Fig.~(\ref{fig:reconstr_1D_finite}) show the field for the long-scale envelope function $f_0$ which is in good agreement with the results from the finite multilayer stack, albeit with some minor discrepancies likely due to the finite extent of the stack and boundary effects not taken into account in our model. 

\begin{figure}[h!]
\subfloat[Exact dispersion diagram (left) and the linear asymptotic approximations (right) obtained by HFH at three points inside the Brillouin zone \eqref{DD_1D_inside} (dashed lines). We also plot the quadratic HFH approximations around $\kappa\delta=0$ and $\kappa\delta=\pi$ \eqref{eq:Omega_final_1D} (plain lines) \label{fig:inside_Sheng_a}]{\includegraphics[trim = 0mm 0mm 0mm 0mm, clip, width = 0.45\linewidth]{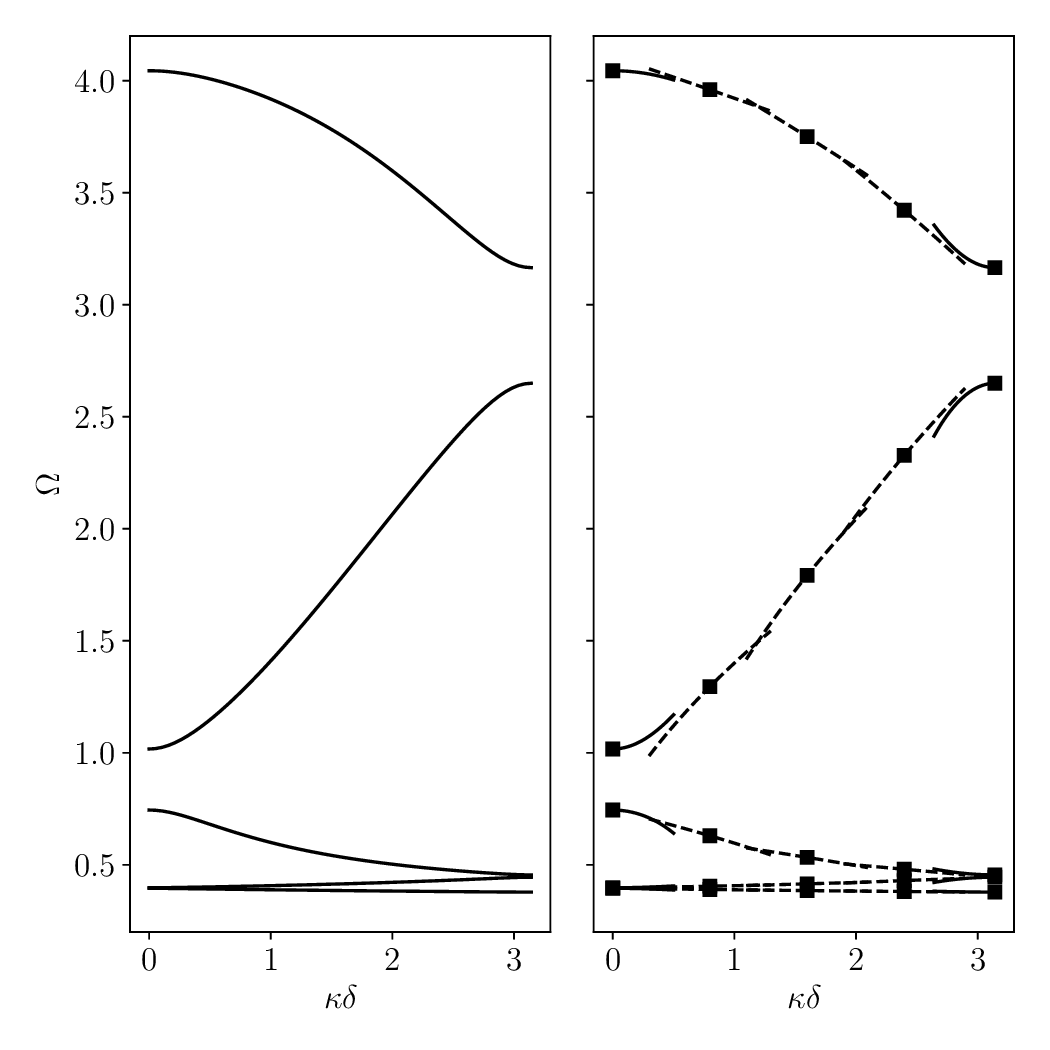}}
\hspace{0.5cm}
\subfloat[Evolution of the coefficient $T_i$ \eqref{T_inside_no_damping} as a function of $\kappa\delta$ for the five modes studied in Figure \ref{fig:inside_Sheng_a}. \label{fig:inside_Sheng_b}]{\includegraphics[trim = 0mm 0mm 0mm 0mm, clip, width = 0.45\linewidth]{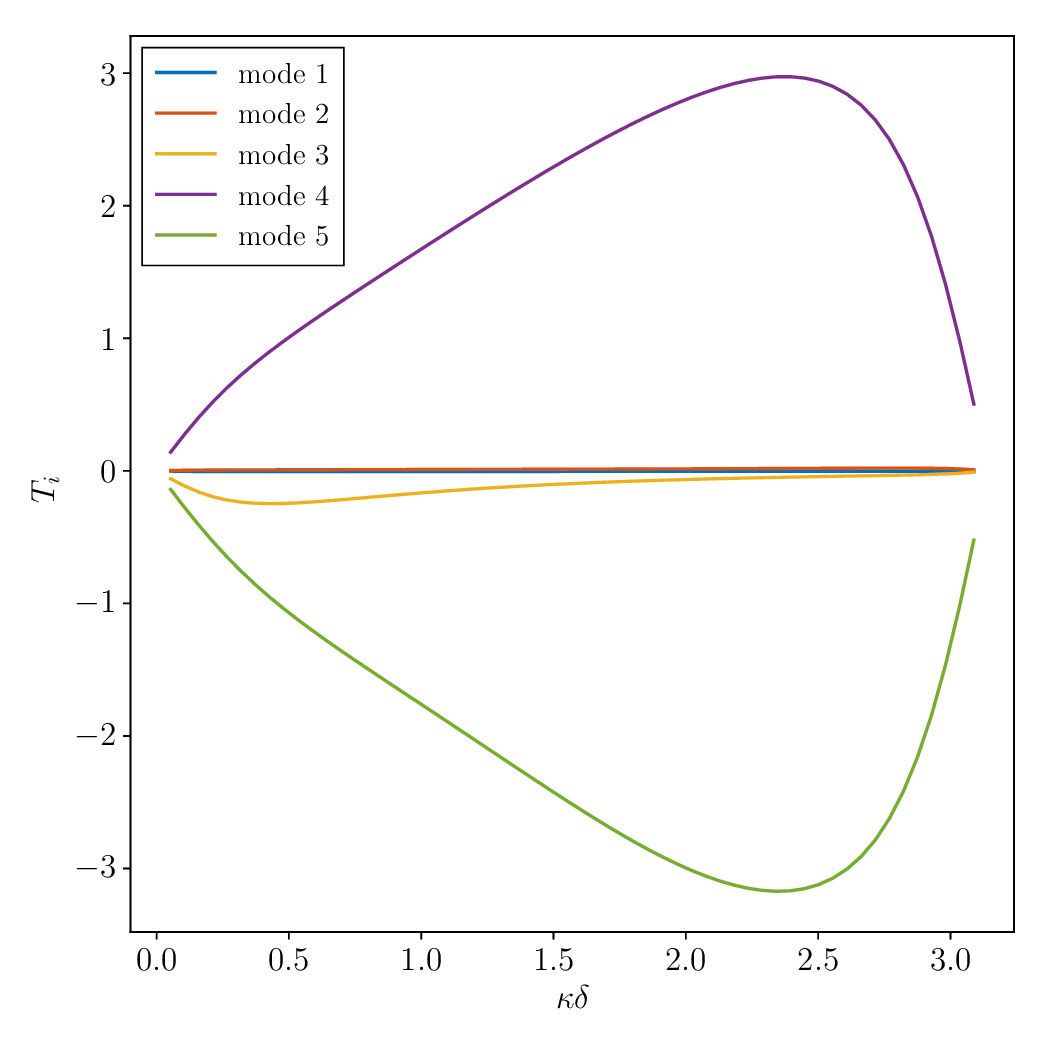}} 
\caption{Effective properties obtained by HFH inside the Brillouin zone (Section \ref{Sec:1D_inside}).}
\label{fig:inside_Sheng}
\end{figure}

\paragraph{Inside the Brillouin zone}
Finally, we make use of the linear asymptotic approximations inside the Brillouin zone, i.e. equation \eqref{DD_1D_inside} of Section \ref{Sec:1D_inside}, on the same example. We see in Figure \ref{fig:inside_Sheng_a} that using this approximation for only three points inside the Brillouin zone and combining it with the quadratic asymptotic approximations at the edges, we almost recover the entire dispersion diagram (obtained with FEM). The effective coefficient $T_i$ also gives an insight on the group velocity in Figure \ref{fig:inside_Sheng_b} since they are proportional. 

\subsubsection{Double eigenvalue case}

\begin{figure}[h!]
\includegraphics[trim = 0mm 0mm 0mm 0mm, clip, width = 1\linewidth]{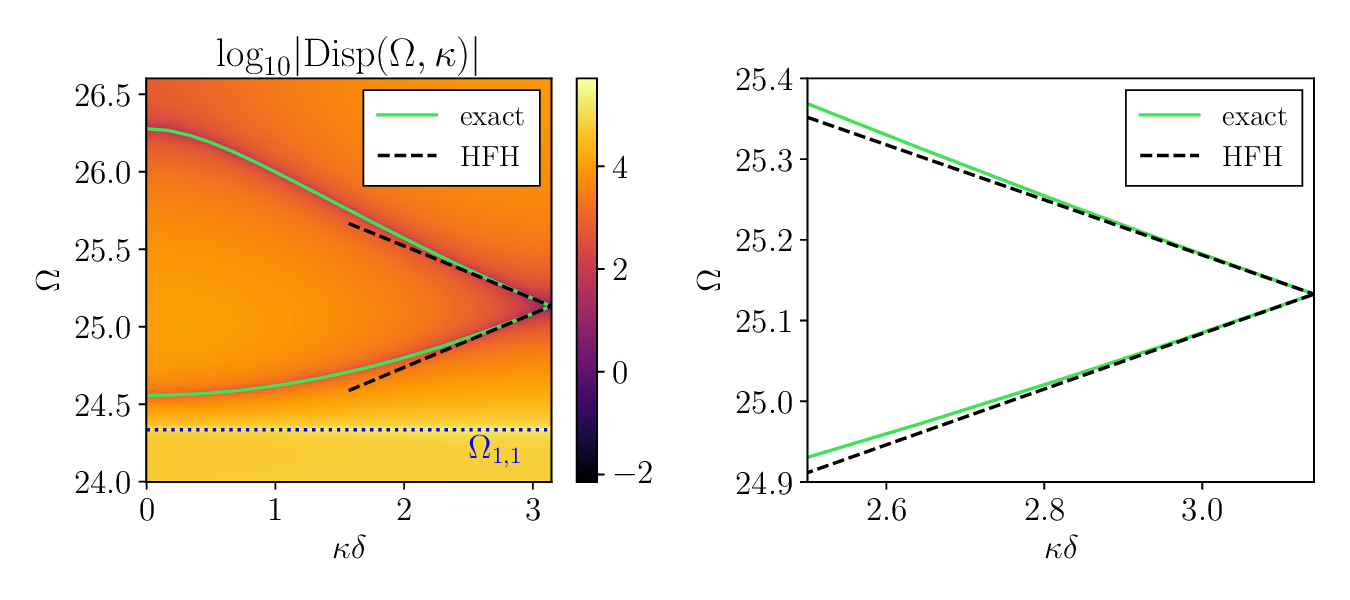} 
\caption{A double eigenvalue case. Comparison of the exact dispersion (green solid lines) with the linear asymptotics obtained by HFH \eqref{Omega_double_1D} (black dashed lines). The colormap on the left panel shows the logarithm of the determinant whose zeros give the dispersion relation, and the dotted horizontal line indicates the position of a pole of $\mathbb{a}$ and of the determinant at $\Omega_{1,1}$. The right panel is a zoom-magnified view close to $\kappa\delta=\pi$. The computed value of the slope is $T_D=17.240$
\label{fig:double_1D}}
\end{figure}

\begin{figure}[h!]
\centering
\includegraphics[trim = 0mm 0mm 0mm 0mm, clip, width = 0.7\linewidth]{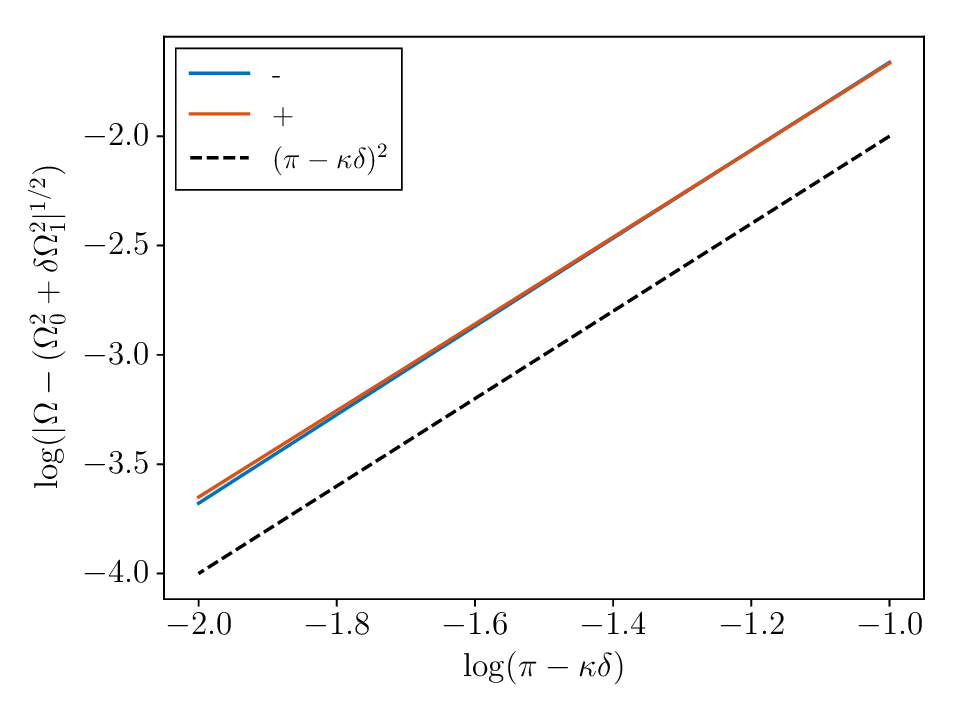} 
\caption{Absolute error near $\kappa\delta=\pi$ between both dispersion relations (the exact dispersion relation solved by FEM and the HFH approximation \eqref{Omega_double_1D}) in a log-log scale (plain lines, blue for the lowest branch ``-'' and red for the upper branch ``+''). The dotted line is the reference order of convergence.
\label{fig:double_1D_convergence}}
\end{figure}

We now investigate a double eigenvalue case (asymptotic approximations developed in Section \ref{Sec:1D_double}). Choosing $\phi=0.5$, $b=1$ and a Drude model with no damping for $a$ with the parameters 
$a_0=a_{\infty}=1$, $\gamma_{1,1} = \Omega_{D,1,1}=0$ and $\Omega_{1,1}=24.3347$ in \eqref{a_adim}. This leads to a double eigenvalue $\Omega_0=25.1322$ at $\kappa\delta=\pi$, with $\mathbb{a}(\Omega_0)\simeq m^2$ where $m$ is an integer ($m=4$ here, cf. \cite{Craster2010}). We note that the value of $\Omega_{1,1}$ is close to $\Omega_0$ and corresponds to a pole of $\mathbb{a}$, meaning the behaviour of the material properties around those frequencies is highly dispersive. Even in this case, our method recovers the expected linear asymptotics with opposite slopes characteristic of a degenerate root, as shown in Figure \ref{fig:double_1D}. 
This is confirmed quantitatively by the curves in Figure \ref{fig:double_1D_convergence} representing the errors between the exact dispersion relation and the linear asymptotic approximations on 
a log-log scale, showing an $\mathcal{O}((\delta\tilde{\kappa})^2)$ convergence for both branches.

\section{Two-dimensional (2D) case}\label{Sec:2D}
We now extend the results to the 2D case; as the method is very similar to the one used in 1D, we will highlight only the differences due to the higher dimensions together with the final asymptotic approximations obtained. 
\subsection{Setting}
\begin{figure}[h!]
\subfloat[2D configuration with zoom on a unit cell]{\includegraphics[trim = 10mm 0mm 0mm 0mm, clip, width = 0.55\linewidth]{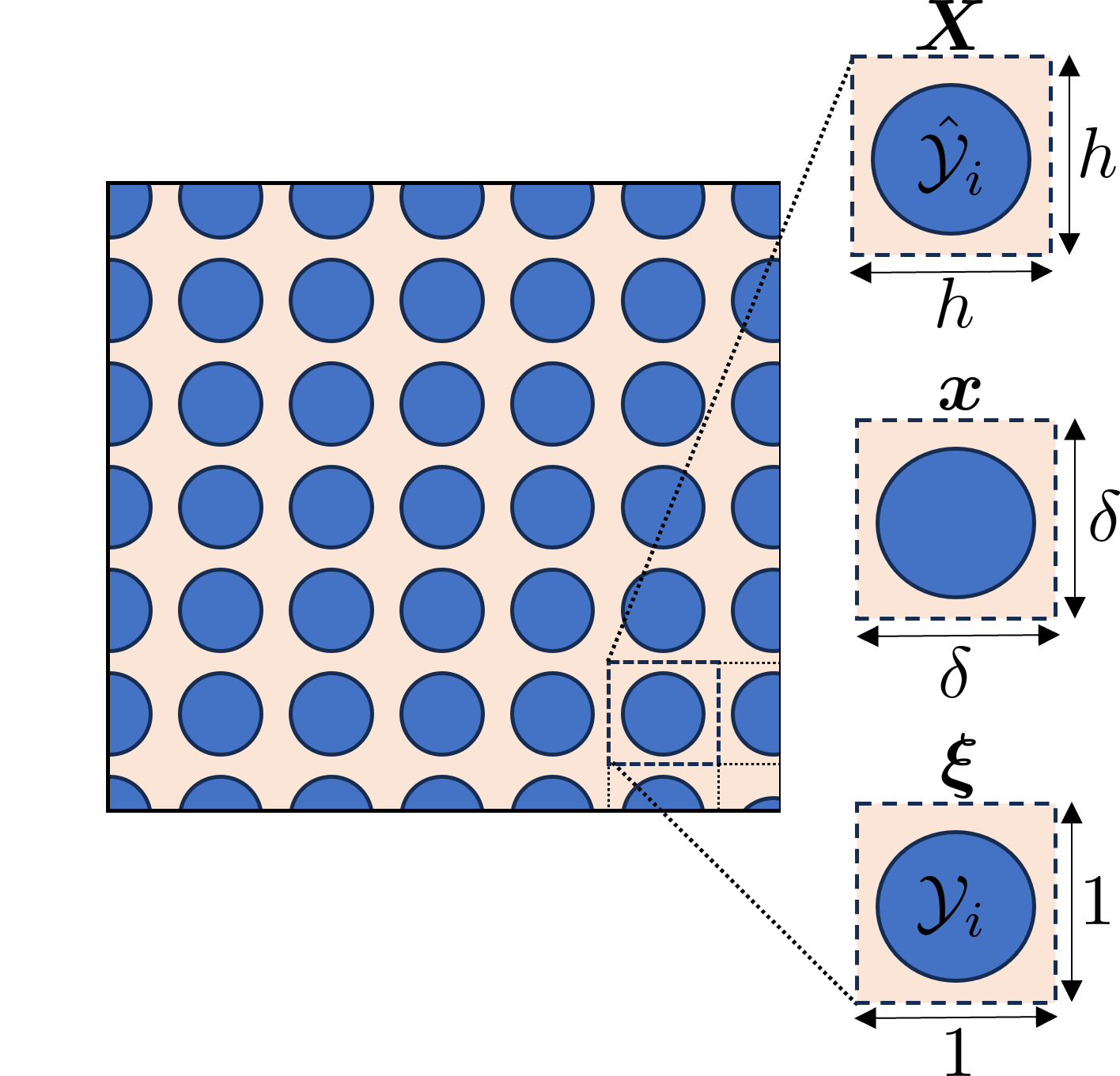}\label{fig:2D_config_a}}
\hspace{0.5cm}
\subfloat[The irreducible
Brillouin zone, in wavenumber space]{\includegraphics[trim = 0mm 0mm 0mm 0mm, clip, width = 0.4\linewidth]{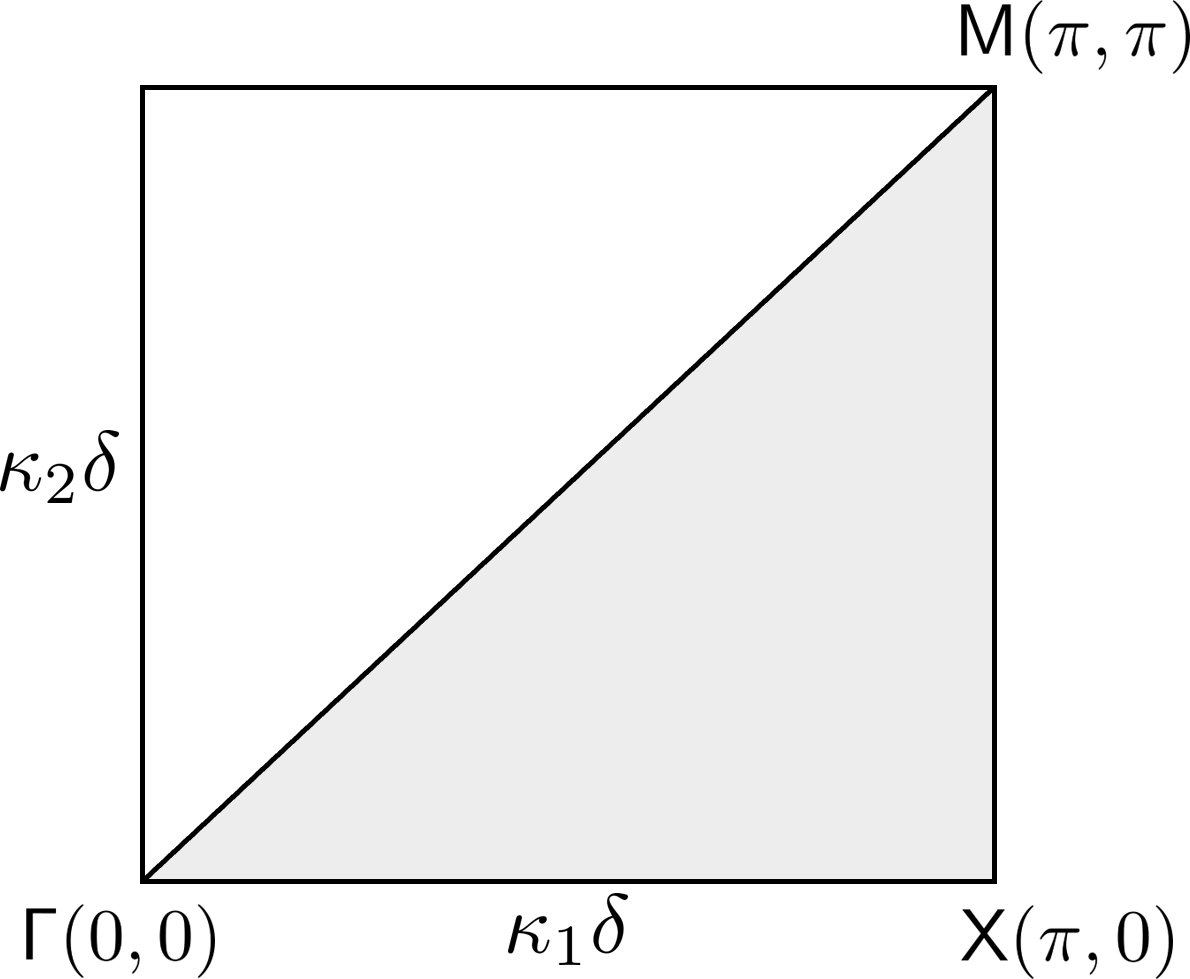}\label{fig:2D_config_b}}
\caption{2D configuration in physical space with zoom on an inclusion in the different coordinate systems (a) and in wavenumber space (b)}
\label{fig:2D_config}
\end{figure}

We consider the Helmholtz equation in a doubly periodic structure on a square lattice of size $h$, see Figure \ref{fig:2D_config_a},
\begin{equation}
    \label{Helmholtz_2D}
    \bm{\nabla}_{\bm{X}}\cdot \left[\hat{a}_h(\bm{X},\omega)\bm{\nabla}_{\bm{X}} U_h(\bm{X}) \right]+\omega^2\hat{ b}_h(\bm{X},\omega)U_h(\bm{X})=0
\end{equation}
The parameters are frequency-dependent inside $\hat{\mathcal{Y}}_i$ in the periodic unit cell, and are simply taken to be constants outside it. As in the 1D case the methodology is developed for any frequency-dependent function and the Drude-Lorentz model is used for numerical examples (see Appendix \ref{Sec:App1} for details on this model). \\
As in 1D, we introduce the two-scales $\bm{x}=\bm{X}/L$ and $\bm{\xi}=\bm{X}/h=\bm{x}/\delta$; we call $\mathcal{Y}$ the unit cell in $\bm{\xi}$-coordinate, with $\mathcal{Y}_i$ the inclusion where the parameters are frequency-dependent. Except for this geometry difference, the non-dimensionalization step is the same as in 1D and a Bloch-Floquet analysis similar to Section \ref{Sec:BlochFloquet} allows us to get the 2D dispersion relation.\\
\subsubsection{Ansatz}
We pick a frequency-wavenumber pair $(\Omega_0,\bm{\kappa})\in \mathbb{C}\times[0,\pi/\delta]^2$ that satisfies the dispersion relation in the irreducible Brillouin zone, see Figure \ref{fig:2D_config_b}. The ansatz for the non-dimensionalized field $u$ and frequency $\Omega$ \eqref{eq:asymptoticrepfield}, together with the expansions for both $a$ and $b$ \eqref{expansion_coefficients} are the same in 2D so we get the following non-dimensionalized equation:
\begin{equation}
\begin{aligned}
  \sum_{j \geqslant 0} &\left[ a(\bm{\xi},\Omega) \left\{  \delta^j  \Delta_{\bm{\xi}} u_j + \delta^{j+1}\left[ \bm{\nabla}_{\bm{\xi}}\cdot \bm{\nabla}_{\bm{x}} u_j+\bm{\nabla}_{\bm{x}}\cdot \bm{\nabla}_{\bm{\xi}} u_j\right] +\delta^{j+2}\Delta_{\bm{x}} u_j  \right\} \right. \\
  &+ \sum_{\ell \geqslant 0}
  \delta^{\ell + j} \Omega_{\ell}^2  b(\bm{\xi},\Omega) u_j ] = 0.
  \label{eq:allinonegoveq_2D}
  \end{aligned}
\end{equation}
\subsubsection{Zeroth-order field}
Collecting the terms of order $\delta^0$, using continuity and periodicity, we get in $\mathcal{Y}$: 
\begin{equation}
    \label{syst_order0_2D}
     \left\lbrace
    \begin{aligned}
   & A_0\Delta_{\bm{\xi}} u_0+\Omega_0^2B_0u_0 = 0 \text{ in } \mathcal{Y} \\
    &u_0 \text{ continuous at } \partial\mathcal{Y}_i \text{ and } \partial\mathcal{Y} \\
    &A_0 \bm{\nabla}_{\bm{\xi}} u_0\cdot\bm{n} \text{ continuous at } \partial\mathcal{Y}_i \text{ and } \partial\mathcal{Y} \\
        & u_0(\bm{x},\xi_1+1,\xi_2)=\mathrm{e}^{\mathrm{i}\kappa_1\delta}u_0(\bm{x},\xi_1,\xi_2) \text{ and } u_0(\bm{x},\xi_1,\xi_2+1)=\mathrm{e}^{\mathrm{i}\kappa_2\delta}u_0(\bm{x},\xi_1,\xi_2)
    \end{aligned}
    \right.   
    \end{equation}

\subsection{High-frequency homogenization for single eigenvalues at the edges of the Brillouin zone}\label{Sec:2D_simple} We start with the case of the edges of the Brillouin zone $\mathsf{\Gamma}$, $\mathsf{X}$ and $\mathsf{M}$, for which $\bm{\kappa}=(0,0),\,(\pi/\delta,0),\,(\pi/\delta,\pi/\delta)$, respectively. 
We choose $\Omega_0$ which is assumed to be a simple eigenvalue associated to the eigenfunction  $\mathcal{U}_0(\bm{\xi},\Omega_0)$. The zeroth-order field then writes 
$$u_0(\bm{x},\bm{\xi})=f_0(\bm{x})\mathcal{U}_0(\bm{\xi},\Omega_0),$$
where $f_0(\bm{x})$ has to be determined. 
\subsubsection{First-order field}
For this single eigenvalue case, we assume that $\Omega_1=0$ and we are looking for the quadratic term $\Omega_2$. Collecting the terms of order $\delta$, we get in $\mathcal{Y}$:
\begin{equation}
    \label{syst_order1_main_2D}
A_0\left(\Delta_{\bm{\xi}}  u_1  +2\bm{\nabla}_{\bm{\xi}}\cdot \bm{\nabla}_{\bm{x}} u_0\right)
+\Omega_0^2 B_0 u_1 =0
\end{equation}
together with:
\begin{itemize}
    \item $u_1(\bm{x},1,\xi_2)=\pm u_1(\bm{x},0,\xi_2)$ and $u_1(\bm{x},\xi_1,1)=\pm u_1(\bm{x},\xi_1,0)$,
    \item continuity for $u_1$ at $\partial\mathcal{Y}_i$ and $\partial\mathcal{Y}$,
    \item continuity for $\left[A_0\left(\bm{\nabla}_{\bm{x}} u_0+\bm{\nabla}_{\bm{\xi}} u_1\right)\right]\cdot\bm{n}$ at $\partial\mathcal{Y}_i$ and $\partial\mathcal{Y}$.
\end{itemize}
Then, we write $u_1$ as: 
\begin{equation}
    \label{form_u1_2D}
    u_1({\bm{x}},\bm{\xi})=f_1({\bm{x}})\mathcal{U}_0({\bm{\xi}})+\bm{\nabla}_{\bm{x}} f_0(\bm{x})\cdot \bm{V}(\bm{\xi})
\end{equation}
where $\bm{V}=(V_1,V_2)$ with for $j=1,2$: 
\begin{equation}
\left\lbrace 
\begin{aligned}
    \label{system_h_2D}
   & A_0\Delta V_j+\Omega_0^2B_0V_j=-2A_0\frac{\partial\mathcal{U}_0}{\partial{{\xi}}_j} \text{ in $\mathcal{Y}$} \\
    &V_j(1,\xi_2)=\pm V_j(0,\xi_2) \text{ and } V_j(\xi_1,1)=\pm V_j(\xi_1,0) \\
    &\text{ continuity for $V_j$ at $\partial\mathcal{Y}_i$ and $\partial\mathcal{Y}$}\\
    &\text{ continuity for  $A_0\left(\bm{\nabla}V_j+\mathcal{U}_0\bm{e}_j\right)\cdot\bm{n}$ at $\partial\mathcal{Y}_i$ and $\partial\mathcal{Y}$. }
    \end{aligned}
\right.
\end{equation}
\subsubsection{Second-order field}
Collecting terms of order $\delta^2$, we get in $\mathcal{Y}$:
\begin{equation}
    \label{syst_order2_main_2D}
A_0\left(\Delta_{\bm{x}} u_0+2\bm{\nabla}_{\bm{\xi}}\cdot \bm{\nabla}_{\bm{x}} u_1+\Delta_{\bm{\xi}} u_2\right)+A_2\Delta_{\bm{\xi}} u_0+\Omega_0^2(B_0u_2+B_2u_0)+\Omega_2^2B_0u_0=0
\end{equation}
together with:
\begin{itemize}
    \item $u_2(\bm{x},1,\xi_2)=\pm u_2(\bm{x},0,\xi_2)$ and $u_2(\bm{x},\xi_1,1)=\pm u_2(\bm{x},\xi_1,0)$
    \item continuity for $u_2$ at $\partial\mathcal{Y}$ and $\partial\mathcal{Y}_i$
    \item continuity for $\left[A_0\left(\bm{\nabla}_{\bm{\xi}} u_2+\bm{\nabla}_{\bm{x}} u_1\right)+A_2\bm{\nabla }_{\bm{\xi}} u_0\right]\cdot\bm{n}$ at $\partial\mathcal{Y}$ and $\partial\mathcal{Y}_i$.
\end{itemize}
We introduce the average operator $\langle\cdot\rangle$ in 2D 
$$\langle f \rangle = \int_{\mathcal{Y}}f(\bm{\xi})\mathrm{d}\bm{\xi}$$
and then consider the expression
$$\langle {u_2}\times\eqref{syst_order0_2D}-u_0\times{\eqref{syst_order2_main_2D}}\rangle)=0.$$
After integration by parts, some algebra and dividing through by $f_0$ we get the final effective equation: 
\begin{equation}
    \label{eq_f0_Omega2_2d}
    T_{ij}\frac{\partial^2f_0}{\partial x_i\partial x_j}+\Omega_2^2f_0=0
\end{equation}
with 
\begin{equation}
    \label{def_Tij}
    T_{ij}=\displaystyle\frac{\left\langle A_0\left( V_i\frac{\partial\mathcal{U}_0}{\partial\xi_j}-\frac{\partial V_i}{\partial\xi_j}\mathcal{U}_0-\mathcal{U}_0^2\delta_{ij}\right)\right\rangle}{ \mathcal{S}(\Omega_0,\mathcal{U}_0,\mathcal{U}_0)}
\end{equation}
where we defined in 2D 
\begin{equation}
\mathcal{S}(\Omega_0,f,g) =  \left\langle\left( b(\cdot,\Omega_0)+\frac{\Omega_0}{2}\frac{\partial  b}{\partial \Omega}(\cdot,\Omega_0)\right)fg - \frac{1}{2\Omega_0}\frac{\partial a}{\partial \Omega}(\cdot,\Omega_0) \bm{\nabla}f\cdot \bm{\nabla}g\right\rangle
    \label{def_S_2D}
\end{equation}
and where we sum over the repeated subscript indexes. \\
From this effective equation, we also get the quadratic term for the dispersion relation 
\begin{equation}
    \label{eq:final_dd_simple_corners_2d}
    \Omega_2^2 =  \tilde{\kappa_i} T_{ij} \tilde{\kappa_j}
\end{equation}
with $\tilde{\kappa}_i=\kappa_i-d_i$ where $d_i=0$ or $\pi/\delta$ depending on the high-symmetry point we choose. \\
The tensor $\bm{T}$ \eqref{def_Tij} encapsulates the effective properties of the periodic structure beyond the quasi-static, classical homogenization, regime. Typically, this tensor may have eigenvalues of markedly different magnitude or of opposite sign, which leads to a change of character of the underlying effective equation, from elliptic to parabolic and from elliptic to hyperbolic, respectively. The former appears at a frequency near band edges and the latter near a frequency at which a saddle point occurs in the corresponding dispersion curves. This has been used notably to design dielectric photonic crystals with spectacular directive emission in the form of + and x wave patterns for a source placed inside in the microwave regime \cite{ceresoli2015dynamic}. The present high-frequency algorithm makes possible the extension of such experiments to the optical wavelengths wherein the periodic assembly of dielectric rods has a frequency dependent refractive index.

\subsection{High-frequency homogenization for repeated eigenvalues at the edges of the Brillouin zone}\label{Sec:2D_repeated}
In this section, we still consider approximation around the edges of the Brillouin zone, but for the case of repeated eigenvalues that gives rise to a linear approximation of the dispersion diagram.
\subsubsection{Zeroth-order field}
System \eqref{syst_order0_2D} still holds but now we assume repeated eigenvalues of multiplicity $N$. We introduce the associated eigenfunctions $\mathcal{U}^{(j)}$ ($j= 1,...,N$).  The solution for the leading-order problem is now
\begin{equation}
    \label{eq:form_u0_double}
    u_0(\bm{x},\bm{\xi}) = \sum_{j=1}^N f_0^{(j)}(\bm{x})\mathcal{U}^{(j)}_0(\bm{\xi}).
\end{equation}
We will denote $\eqref{syst_order0_2D}^{(l)}$ the system $\eqref{syst_order0_2D}$ satisfied by $u_0 ^{(l)}=f_0^{(l)}\mathcal{U}^{(l)}_0$ for $l\in\lbrace 1,...,N\rbrace$.
\subsubsection{First-order field}
The main difference is that now $\Omega_1 \neq 0$, therefore the system for the first-order field is modified and collecting the terms of order $\delta$ we now get in $\mathcal{Y}$:
\begin{equation}
    \label{syst_order1_main_2D_double}
A_0\left(\Delta_\xi  u_1  +2\bm{\nabla}_\xi\cdot \bm{\nabla}_x u_0\right)
+\Omega_0^2 (B_0 u_1+B_1 u_0) + \Omega_1^2 B_0 u_0 +A_1 \Delta_\xi u_0 =0
\end{equation}
together with:
\begin{itemize}
    \item  $u_1(\bm{x},1,\xi_2)=\pm u_1(\bm{x},0,\xi_2)$ and $u_1(\bm{x},\xi_1,1)=\pm u_1(\bm{x},\xi_1,0)$,
    \item continuity for $u_1$ at $\partial\mathcal{Y}_i$ and $\partial\mathcal{Y}$,
    \item continuity for $\left[A_0\left(\bm{\nabla}_x u_0+\bm{\nabla}_\xi u_1\right)\right]\cdot\bm{n}+A_1\bm{\nabla}_\xi u_0 \cdot\bm{n}$ at $\partial\mathcal{Y}_i$ and $\partial\mathcal{Y}$.
\end{itemize}
Let us pick one $l\in\lbrace 1,...,N\rbrace$ and compute $\langle \eqref{syst_order0_2D}^{(l)} \times u_1 - \eqref{syst_order1_main_2D_double} \times u_0 ^{(l)} \rangle $. By integrating by parts, using different continuity conditions and dividing through by $f_0^{(l)}$ we get the effective equation: 
\begin{equation}
    \sum_{j\neq l}\langle A_0 \bm{W}_0^{jl} \rangle \cdot \bm{\nabla}_{\bm{x}} f_0^{(j)} = -\Omega_1^2 \sum_j \mathcal{S}(\Omega_0,\mathcal{U}_0^{(l)},\mathcal{U}_0^{(j)}) f_0^{(j)} 
\end{equation}
with 
\begin{equation}
    \label{W_0_double_2D}
    \bm{W}_0^{jl} = \mathcal{U}_0^{(l)}\bm{\nabla}_{\bm{\xi}}\mathcal{U}_0^{(j)}- \mathcal{U}_0^{(j)}\bm{\nabla}_{\bm{\xi}}\mathcal{U}_0^{(l)}.
\end{equation}
We set $f_0^{(l)}=\hat{f}_0^{(l)}\mathrm{exp}(i\tilde{\kappa_j} x_j)$ and get the following system of equations:
\begin{equation}
    \label{eq_matrix_system_eq_double}
    \bm{C} \hat{\bm{F}}_0 = \bm{0}
\end{equation}
with the $2\times2$ matrix $\bm{C}$ defined by 
\begin{equation}
    \label{Cjk_double}
    C_{lj} = \mathrm{i} \langle  a(\cdot,\Omega_0) \bm{W}_0^{jl}\rangle \cdot \tilde{\bm{\kappa}}+\Omega_1^2   \mathcal{S}(\Omega_0,\mathcal{U}_0^{(l)},\mathcal{U}_0^{(j)})  .
\end{equation}
The value of $\Omega_1^2$ is then obtained by solving $\mathrm{det}(\bm{C})=0$.

\begin{remark}
    For double eigenvalues we get the following expression for the linear term (opposite slopes)
    \begin{equation}
        \label{analytique_final_double}
\Omega_1^2  = \pm\frac{\tilde{\bm{\kappa}}\cdot \langle a(\cdot,\Omega_0)  \bm{W_0}^{12}\rangle }{\left(
    \mathcal{S}(\Omega_0,\mathcal{U}_0^{(1)},\mathcal{U}_0^{(1)}) 
    \mathcal{S}(\Omega_0,\mathcal{U}_0^{(2)},\mathcal{U}_0^{(2)}) 
    -(\mathcal{S}(\Omega_0,\mathcal{U}_0^{(1)},\mathcal{U}_0^{(2)})) ^2
    \right)^{1/2}}.
    \end{equation}
   
\end{remark}

\subsection{High-frequency homogenization for nearby eigenvalues at the \\edges of the Brillouin zone\label{sec:nearby2D}}
In this section, we again consider asymptotic approximations around the edges of the Brillouin zone, and we assume that the eigenvalues are single but close to each other. More precisely, we consider $N$ eigenvalues close to each other so that the distances between them scale into the small parameter $\delta$ and write
\begin{equation}
    \label{eq:distance_ev_nearby_2D}
    \delta\alpha_l=(\Omega_0^{(l)})^2-(\Omega_0^{(1)})^2,
\end{equation}
for $l\in\lbrace 1,...,N\rbrace$. To take into account their competitive nature, we assume that the leading order field is
\begin{equation}
    u_0(\bm{x},\bm{\xi})=\sum_{j=1}^N f_0^{(j)}(\bm{x})\mathcal{U}_0^{(j)}(\bm{\xi}).
\end{equation}
As in 1D, the ansatz is considered around the eigenvalue $\Omega_0^{(1)}.$ \\
In that case, the residual term for the zeroth order equation is 
\begin{equation}
    \label{residual_2D}
    \delta\sum_{q\neq 1}\alpha_q u_0^{(q)}\left[\frac{1}{2A_0^{(1)}}\Omega_0^{(1)}B_0^{(1)}\frac{\partial a}{\partial\Omega}(\cdot,\Omega_0^{(1)})-B_0^{(1)}-\frac{1}{2}\Omega_0^{(1)}\frac{\partial b}{\partial\Omega}(\cdot,\Omega_0^{(1)})\right]
\end{equation}
that will in turn modify the equation for the first order field 
in $\mathcal{Y}$ to be:
\begin{equation}
    \label{syst_order1_main_2D_nearby}
    \begin{aligned}
&A_0^{(1)}\left(\Delta_\xi  u_1  +2\bm{\nabla}_\xi\cdot \bm{\nabla}_x u_0\right)
+(\Omega_0^{(1)})^2 (B_0^{(1)} u_1+B_1 u_0) +\Omega_1^2 B_0^{(1)} u_0 +A_1 \Delta_\xi u_0 \\
&+\sum_{q\neq 1}\alpha_q u_0^{(q)}\left[\frac{1}{2A_0^{(1)}}\Omega_0^{(1)}B_0^{(1)}\frac{\partial a}{\partial\Omega}(\cdot,\Omega_0^{(1)})-B_0^{(1)}-\frac{1}{2}\Omega_0^{(1)}\frac{\partial b}{\partial\Omega}(\cdot,\Omega_0^{(1)})\right]=0,
\end{aligned}
\end{equation}
together with $u_1(\bm{x},1,\xi_2)=\pm u_1(\bm{x},0,\xi_2)$ and $u_1(\bm{x},\xi_1,1)=\pm u_1(\bm{x},\xi_1,0)$, continuity for $u_1$ at $\partial\mathcal{Y}_i$ and $\partial\mathcal{Y}$, and continuity for $\left[A_0\left(\bm{\nabla}_x u_0+\bm{\nabla}_\xi u_1\right)\right]\cdot\bm{n}+A_1\bm{\nabla}_\xi u_0 \cdot\bm{n}$ at $\partial\mathcal{Y}_i$ and $\partial\mathcal{Y}$.

Let us pick one $l\in\lbrace 1,...,N\rbrace$ and consider $\langle \eqref{syst_order0_2D}^{(l)} \times u_1 - \eqref{syst_order1_main_2D_nearby} \times u_0 ^{(l)} \rangle =0$. By integrating by parts, using the different continuity conditions, dividing through by $f_0^{(l)}$, and neglecting the higher-order terms we get the effective equation
\begin{equation}
    \sum_{j=1}^N\langle A_0 \bm{W}_\mathrm{nby}^{jl} \rangle \cdot \bm{\nabla}_x f_0^{(j)} = \sum_{j=1}^N \lbrace -\Omega_1^2 \langle\mathcal{S}(\Omega_0^{(1)},\mathcal{U}_0^{(l)},\mathcal{U}_0^{(j)})\rangle+\alpha_j \mathcal{N}^{jl}  \rbrace f_0^{(j)} 
\end{equation}
with 
\begin{equation}
    \label{W_0_nearby_2D}
    \left\lbrace
    \begin{aligned}
        &\bm{W}_\mathrm{nby}^{jl} = \mathcal{U}_0^{(l)}\bm{\nabla}_\xi\mathcal{U}_0^{(j)}- \mathcal{U}_0^{(j)}\bm{\nabla}_\xi\mathcal{U}_0^{(l)}, \\
     &\mathcal{N}^{jl}=\left\langle b(\cdot,\Omega_0^{(1)})\left( 1 + \frac{\Omega_0^{(1)}}{2} \frac{\partial }{\partial\Omega}[\log (b/a)](\cdot,\Omega_0^{(1)}) \right)\mathcal{U}_0^{(j)}\mathcal{U}_0^{(l)}\right\rangle.
    \end{aligned}
    \right.
\end{equation}

The linear term $\Omega_1$ of the dispersion relation is given by solving $\mathrm{det}(\bm{C}_\mathrm{nby})=0$ where the $N\times N$ matrix $\bm{C}_\mathrm{nby}$ is defined by 
\begin{equation}
  (\bm{C}_\mathrm{nby})_{lj}  = \mathrm{i}\left\langle a(\cdot,\Omega_0^{(1)})  \bm{W}_\mathrm{nby}^{jl} \right\rangle \cdot\tilde{\bm{\kappa}}+\Omega_1^2\mathcal{S}(\Omega_0^{(1)},\mathcal{U}_0^{(l)},\mathcal{U}_0^{(j)})-\alpha_j\mathcal{N}^{jl}.
  \label{eq:nearby_2D_det}
\end{equation}
We can notice that we recover the repeated eigenvalues case when the distances $\alpha_j$ tend to 0.
\begin{remark}
    In the case of two nearby eigenvalues the dispersion relation is given by 
    \begin{equation}
    \begin{aligned}
        \label{Omega1_2nearby_2D}
        &\Omega_1^4\left[\mathcal{S}(\Omega_0^{(1)},\mathcal{U}_0^{(1)},\mathcal{U}_0^{(1)})\mathcal{S}(\Omega_0^{(1)},\mathcal{U}_0^{(2)},\mathcal{U}_0^{(2)})-(\mathcal{S}(\Omega_0^{(1)},\mathcal{U}_0^{(1)},\mathcal{U}_0^{(2)}))^2 \right]\\
        & +\Omega_1^2\alpha_2\left[-\mathcal{N}^{22}\mathcal{S}(\Omega_0^{(1)},\mathcal{U}_0^{(1)},\mathcal{U}_0^{(1)})+\mathcal{N}^{21}\mathcal{S}(\Omega_0^{(1)},\mathcal{U}_0^{(1)},\mathcal{U}_0^{(2)}) \right] \\
        & -\left(\langle a(\cdot,\Omega_0^{(1)}) \bm{W}_\mathrm{nby}^{21}\rangle\cdot\tilde{\bm{\kappa}}\right)^2-\mathrm{i}\alpha_2\mathcal{N}^{21}\langle a(\cdot,\Omega_0^{(1)}) \bm{W}_\mathrm{nby}^{21}\rangle\cdot\tilde{\bm{\kappa}}=0.
        \end{aligned}
    \end{equation}
\end{remark}

\subsection{High-frequency homogenization for simple eigenvalues inside the Brillouin zone (no damping)}\label{Sec:2D_inside}
Then, when no damping is considered, we are able to get linear asymptotic approximations near $\bm{\kappa}^\star$ which is not one of the high-symmetry points. The method being very similar to the 1D case, we only give the effective equation 
\begin{equation}
    \label{eq_f0_int_2D}
    \bm{T}^\mathrm{int}\cdot \bm{\nabla}f_0-\mathrm{i}\Omega_1^2f_0=0
\end{equation}
where $\bm{T}^\mathrm{int}=(T_1^\mathrm{int},T_2^\mathrm{int})$ with $T^\mathrm{int}_{i}$ given, for $i=1,2$ by:
\begin{equation}
    \label{T_int_2D}
    T^\mathrm{int}_{i}=\frac{2\langle A_0 \Im(\frac{\partial \mathcal{U}_0}{\partial\xi_i}\bar{\mathcal{U}}_0)\rangle}{\mathcal{S}(\Omega_0,\mathcal{U}_0,\bar{\mathcal{U}_0}) }.
\end{equation}
And consequently, the dispersion relation reads
\begin{equation}
    \label{DD_2D_inside}
    \Omega_1^2=\bm{T}^\mathrm{int}\cdot(\bm{\kappa}-\bm{\kappa}^\star).
\end{equation}
 \subsection{Low-frequency case}
  \label{eq:asymptoticrepfield}
  Finally, we also obtain the classical low-frequency homogenized equation. In that case, we consider $\Omega_0=0$ in \eqref{syst_order0_2D}, which leads to the fact that $\mathcal{U}_0$ is uniform, say $\mathcal{U}_0=1$. Then, we write $u_1(x,\xi)=\bm{\nabla}_{\bm{x}} f_0(\bm{x})\cdot \bm{V}(\bm{\xi})$ where $u_1$ satisfies \eqref{system_h_2D} with $\Omega_0=0$ and $u_0=f_0$.  Therefore $\bm{V}$ satisfies \eqref{system_h_2D} with $\Omega_0=0$ and $\mathcal{U}_0=1$. Integrating \eqref{syst_order2_main_2D} on a unit cell for $\Omega_0=0$ then leads to the usual homogenized equation
\begin{equation}
    \label{eq:LF_limit_2D}
    \left\langle A_0\left(\delta_{ij}+\frac{\partial}{\partial \xi_i}V_j\right)\right\rangle\frac{\partial^2}{\partial x_i\partial x_j}f_0(x)+\Omega_2^2\langle B_0\rangle f_0(x) = 0
\end{equation}
together with the dispersion relation 
\begin{equation}
    \label{eq:LF_limit_2D_disp}
    \Omega_2^2=\left\langle A_0\left(\delta_{ij}+\frac{\partial}{\partial \xi_i}V_j\right) \right\rangle \langle B_0\rangle^{-1}\kappa_{i}\kappa_j .
\end{equation}
It is well-known one can identify the homogenized matrix in (\ref{eq:LF_limit_2D}) making use of the acoustic band near the origin \cite{Bensoussan2011,Conca1997}, albeit for non-dispersive media (see section 3.2 in  \cite{Conca1997} for a summary of results published back in 1978 in the first edition of \cite{Bensoussan2011}).

\subsection{Numerical example}

We consider here a two-dimensional lattice of dispersive rods as studied by Br{\^u}l{\'e} et al. \cite{Brule2016}. 
The material parameters are the permittivity $\varepsilon$ given by a single resonance Drude model $\varepsilon(\Omega) = 1- \Omega_{p}^2/(\Omega(\Omega+{\rm i }\gamma))$, with 
$\Omega_{p}/2\pi= 1.1$,  $\gamma/2\pi= 0.05$, 
and non magnetic material with permeability $\mu=1$. 
In the TM polarization (s-polarization) case this corresponds in our notations to $\mathbb{a}=1/\mu$ and $\mathbb{b} =\varepsilon$, whereas in the TE polarization (p-polarization) case $\mathbb{a}=1/\varepsilon$ and $\mathbb{b} =\mu$. The dispersion diagrams are computed by FEM from the exact dispersion function and here again will be denoted as the \textit{exact dispersion diagrams} by opposition to the asymptotic approximations obtained by HFH. 

\begin{figure}[h!]
\centering
\subfloat[Real part]{\includegraphics[trim = 0mm 0mm 0mm 0mm, clip, width = 0.33\linewidth]{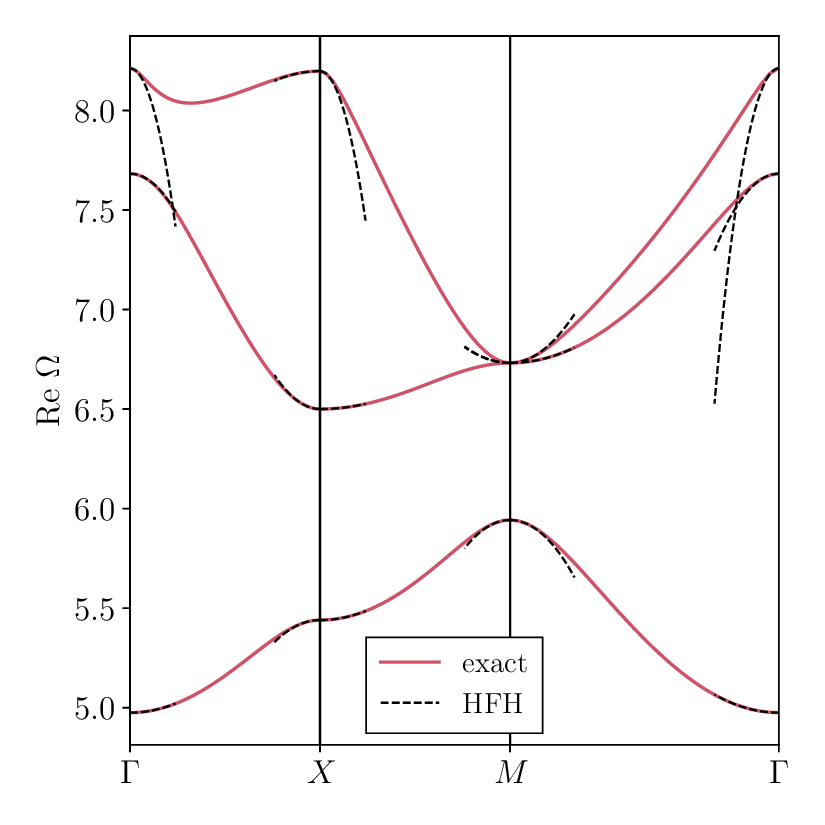}}
\subfloat[Imaginary part]{\includegraphics[trim = 0mm 0mm 0mm 0mm, clip, width = 0.33\linewidth]{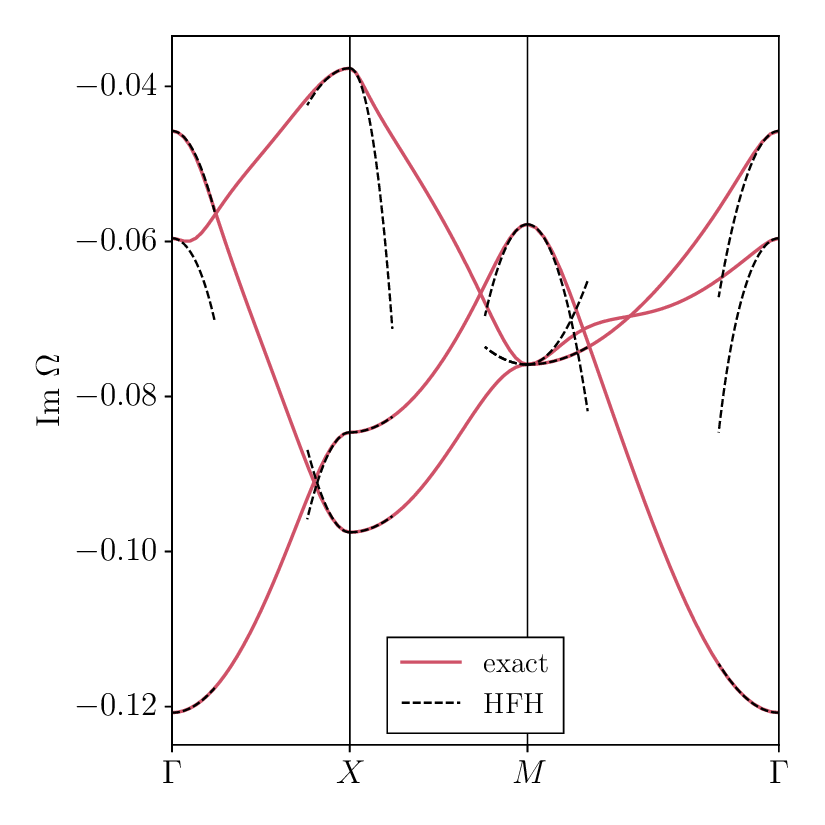}}
\subfloat[Complex plane]{\includegraphics[trim = 0mm 0mm 0mm 0mm, clip, width = 0.33\linewidth]{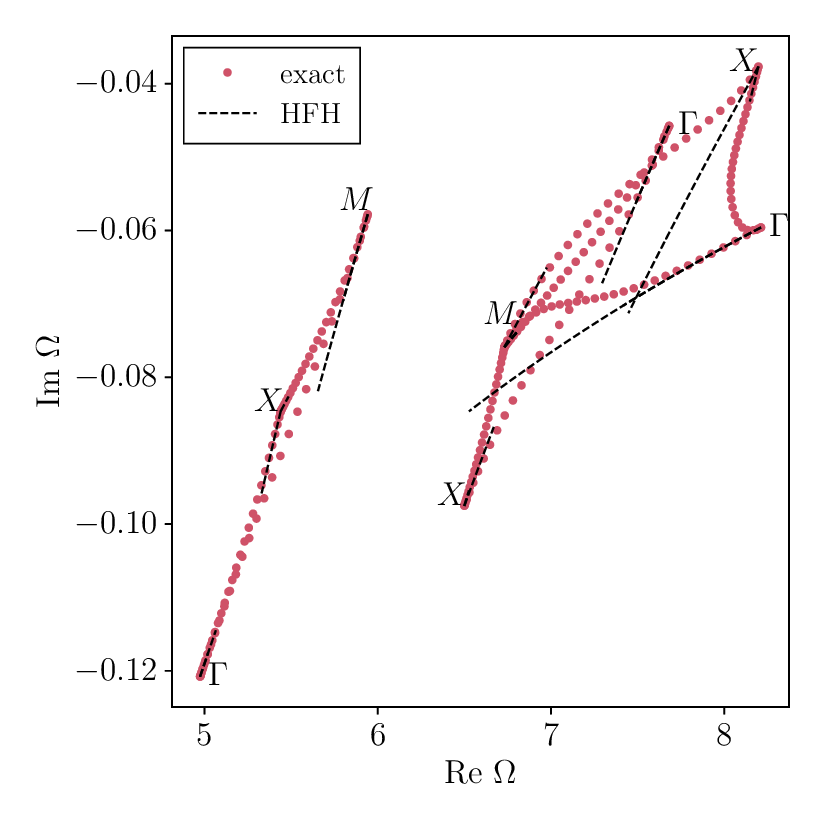}}
\caption{Simple eigenvalue HFH approximations \eqref{eq:final_dd_simple_corners_2d} (dashed lines) for the first three bands of the square photonic crystal in TM polarization.  }
\label{fig:band_3_2d_phc_square}
\end{figure}

\paragraph{Single eigenvalues at symmetry points}
The first example is a square array of period $h$ of square rods of size $L=0.806 h$
made of this Drude permittivity in vacuum. Assuming first TM polarization, 
we plot the dispersion along the edge of the first Brillouin zone on Figure \ref{fig:band_3_2d_phc_square} for the first three modes. The results obtained by HFH in Section \ref{Sec:2D_simple}
approximate well the dispersion behaviour locally around the symmetry points. 
The spectral features showing deformed triangles in the complex plane and including an 
intertwining of the second and third band is well recovered by the HFH asymptotic approximations 
(see panel (c) in Figure \ref{fig:band_3_2d_phc_square}).

\paragraph{Nearby eigenvalues}

\begin{figure}[h!]
\subfloat[Real part for a cluster of four eigenvalues around $\mathsf{\Gamma}$ point.\label{fig:nearby_2d_real}]{\includegraphics[trim = 0mm 0mm 0mm 0mm, clip, width = 0.45\linewidth]{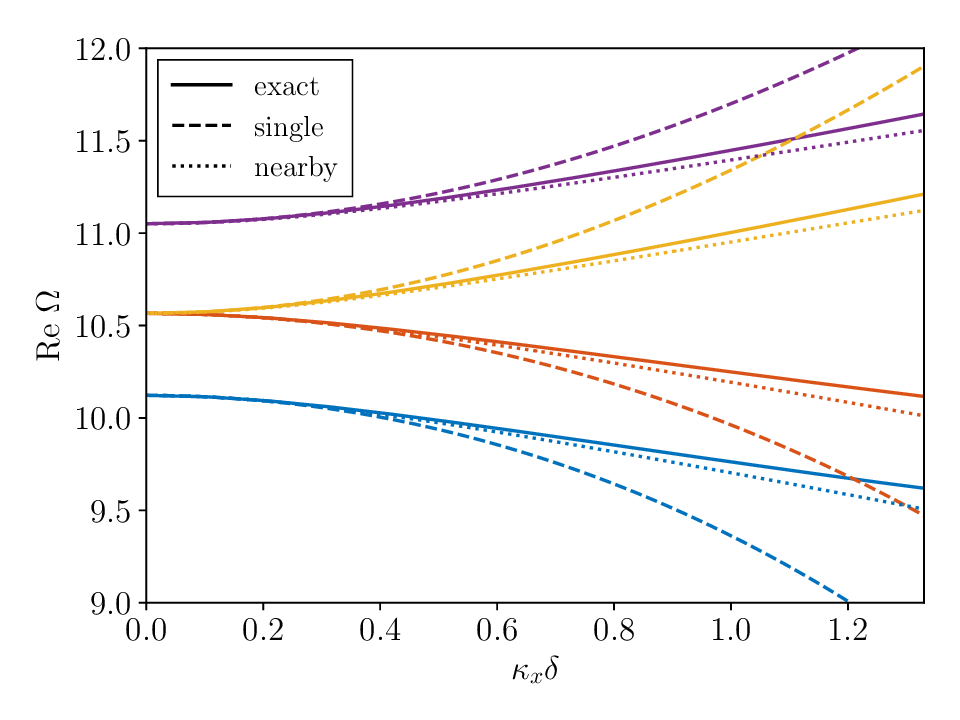}}
\hspace{0.5cm}
\subfloat[Imaginary part for a cluster of four eigenvalues around $\mathsf{\Gamma}$ point.\label{fig:nearby_2d_imag}]{\includegraphics[trim = 0mm 0mm 0mm 0mm, clip, width = 0.45\linewidth]{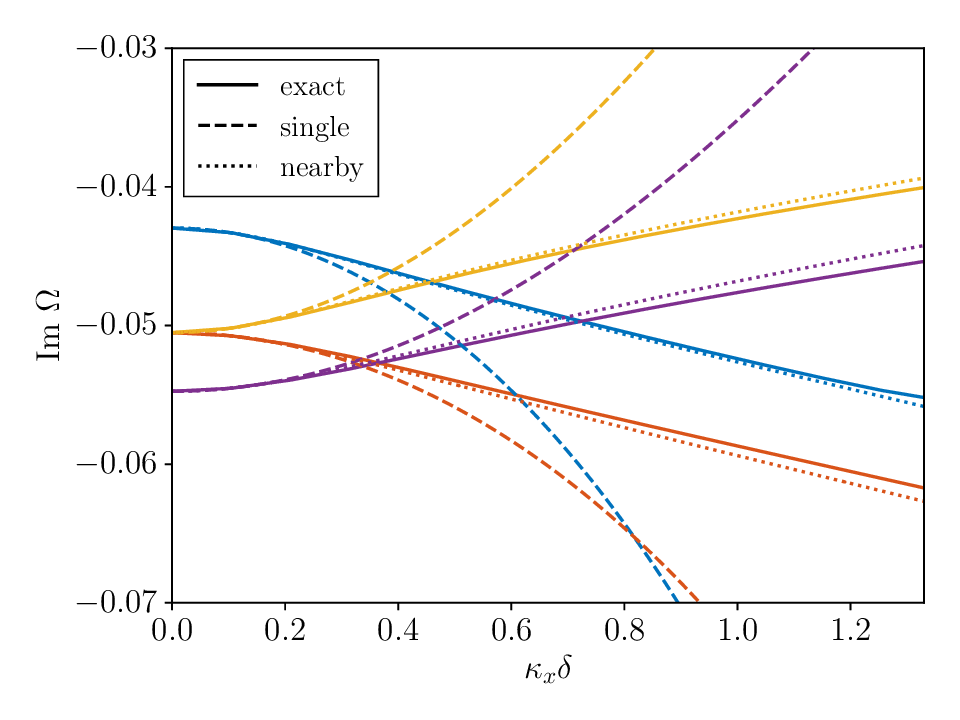}}
\caption{Comparison of the exact dispersion relation approximated by FEM (solid lines) and the effective one obtained by HFH for the single eigenvalue case \eqref{eq:final_dd_simple_corners_2d} (dashed lines) and with nearby approximations \eqref{eq:nearby_2D_det} (dotted lines).}
\label{fig:nearby_2d}
\end{figure}

\begin{figure}[h!]
\centering
\includegraphics[trim = 0mm 0mm 0mm 0mm, clip, width = 0.9\linewidth]{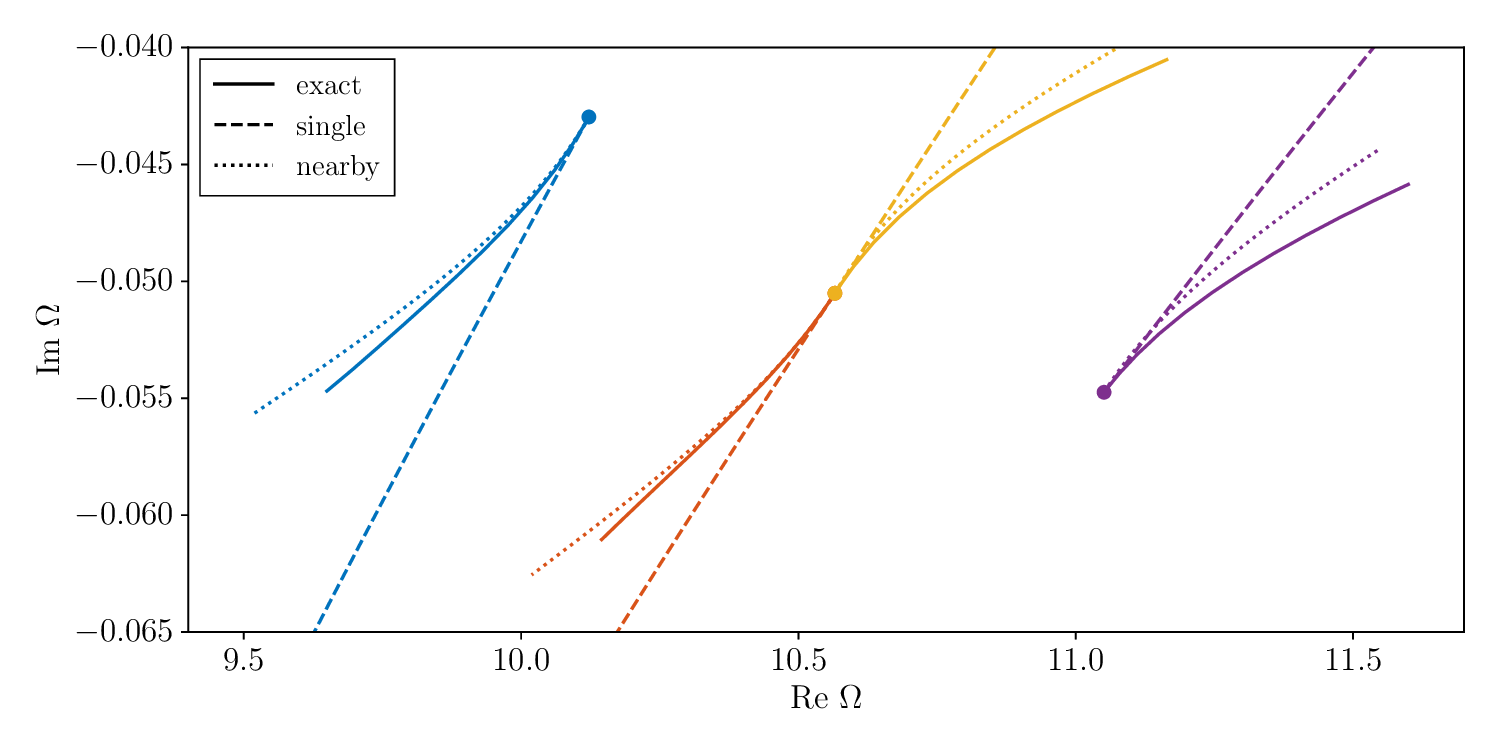} 
\caption{Comparison in the complex plane of the exact dispersion relation computed by FEM (solid lines) and the effective one obtained by HFH for the single eigenvalue case \eqref{eq:final_dd_simple_corners_2d} (dashed lines) and with nearby approximations \eqref{eq:nearby_2D_det} (dotted lines) around $\mathsf{\Gamma}$ point. The circle markers correspond to $\kappa_x\delta=0$.}
\label{fig:nearby_2d_complex}
\end{figure}

We now make use of the linear asymptotic approximations of Section \ref{sec:nearby2D}  for single eigenvalues near edges of the Brillouin zone. Our focus is the cluster of four eigenfrequencies along $\mathsf{\Gamma} \mathsf{X}$ close to the point $\mathsf{\Gamma}$: the results of FEM computations are displayed on Fig.~(\ref{fig:nearby_2d}). 
The exact dispersion curves (solid lines) are correctly approximated by the single 
eigenvalue HFH model (dashed curves), and even better so by the nearby case (dotted lines). Indeed the nearby approximation leads to a better prediction of the local behaviour of bands, 
and as in the 1D numerical example this is particularly striking for the imaginary parts. 
The dispersion diagram in the complex plane  and the corresponding asymptotic approximations are reported in Fig.~(\ref{fig:nearby_2d_complex}).

\paragraph{Simple eigenvalues inside the Brillouin zone}

\begin{figure}[h!]
\subfloat[TE polarization]{\includegraphics[trim = 0mm 0mm 0mm 0mm, clip, width = 0.45\linewidth]{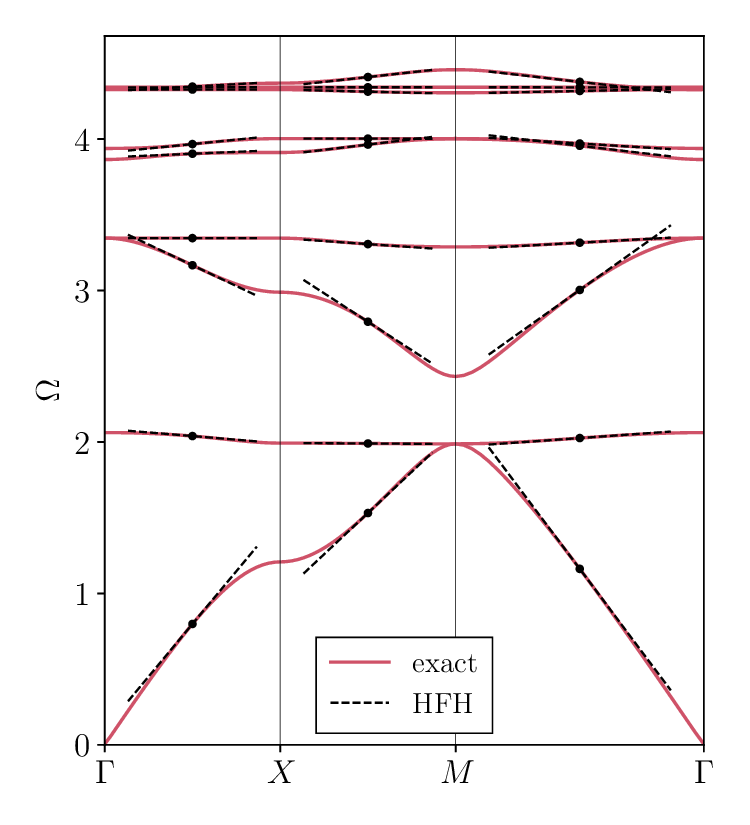}}
\hspace{0.5cm}
\subfloat[TM polarization]{\includegraphics[trim = 0mm 0mm 0mm 0mm, clip, width = 0.45\linewidth]{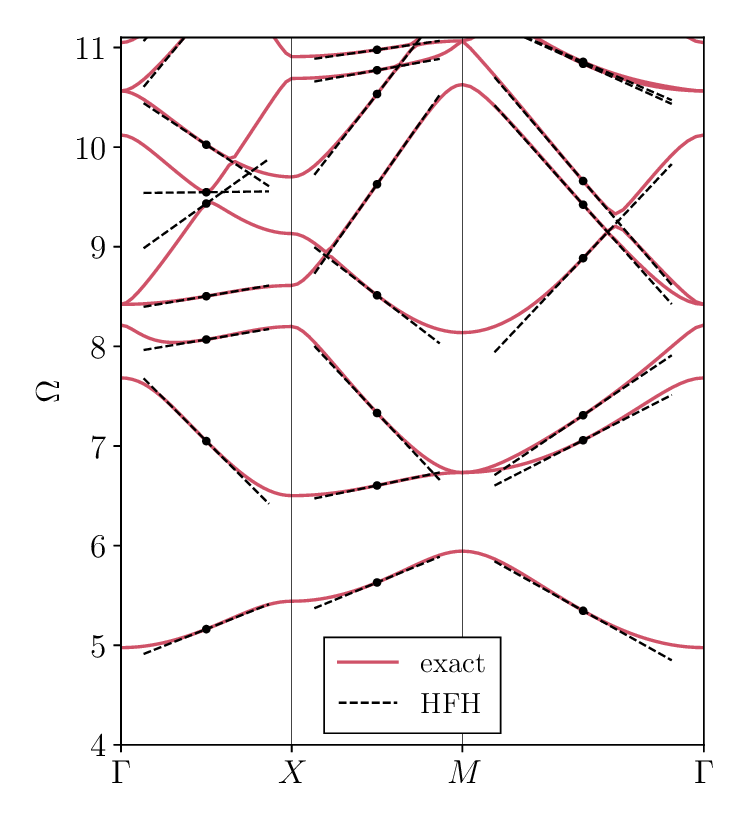}}
\caption{Superposition of the exact dispersion diagram (plain lines) and the asymptotic approximations obtained by HFH (dashed lines) for points inside the Brillouin zone \eqref{DD_2D_inside}. }
\label{fig:inside_no_damping_2D}
\end{figure}

For eigenvalues not located at symmetry points, we study the same structure but setting $\gamma=0$. Making use of the linear approximation, (\ref{DD_2D_inside}), obtained in Section \ref{Sec:2D_inside}, 
we recover locally the behaviour of the bands (see Figure \ref{fig:inside_no_damping_2D}).

\paragraph{Repeated eigenvalues}

\begin{figure}[h!]
\centering
\includegraphics[trim = 0mm 0mm 0mm 0mm, clip, width = 0.6\linewidth]{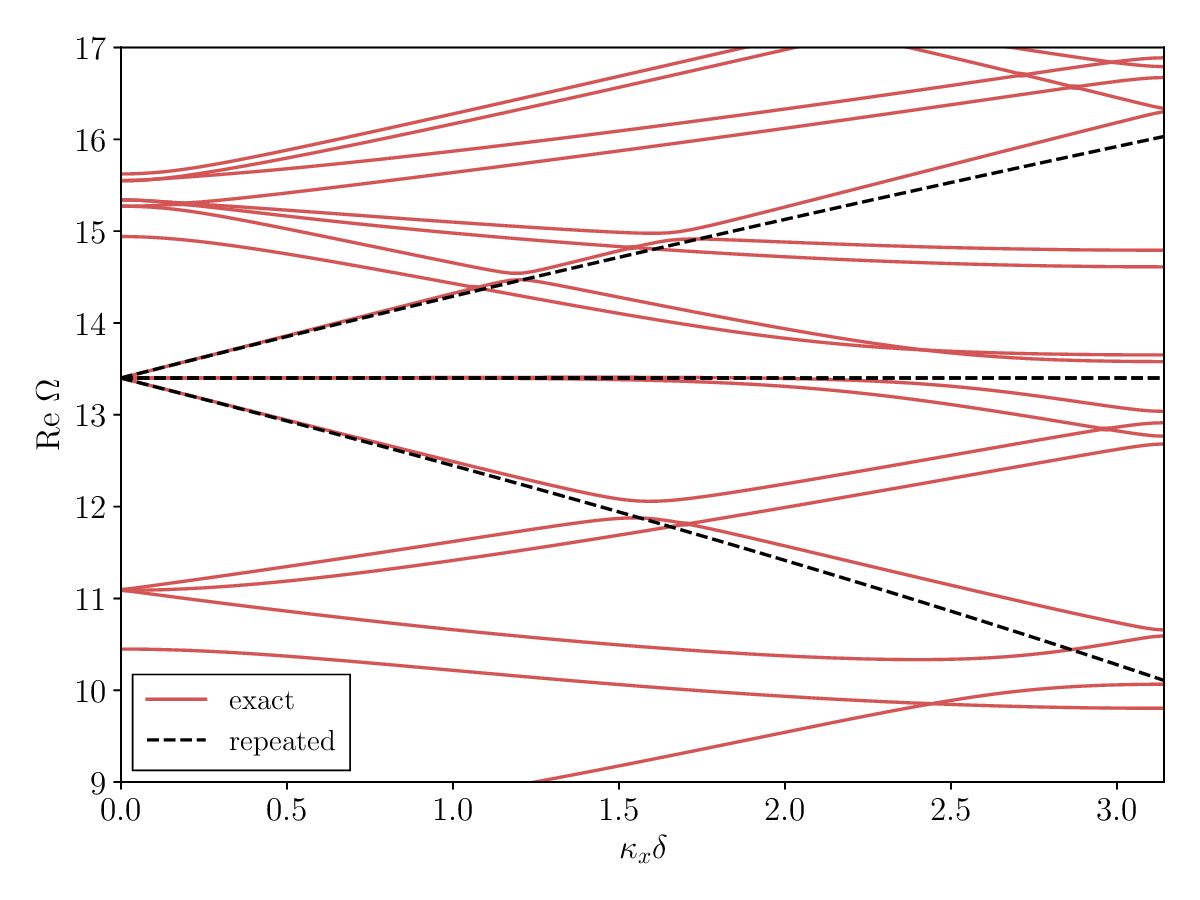} 
\caption{Comparison of the exact dispersion relation (red solid lines) and the effective one obtained by HFH for the repeated eigenvalue case \eqref{Cjk_double} (black dashed lines) around $\mathsf{\Gamma}$ point. }
\label{fig:phc2D_repeated}
\end{figure}

For certain choices of the dispersive behaviour and material distribution, there exist accidental degeneracies. In this section we choose circular inclusions with $R=0.364 h$ and use the same Drude model for the permittivity with $\Omega_{p}/2\pi= 1.434$ and $\gamma = 0$. The band structure in TM polarization is represented on Figure \ref{fig:phc2D_repeated}, where we 
can see around $\mathsf{\Gamma}$ point the coalescence of four bands around $\Omega = 13.4$. 
The repeated eigenvalue approximation obtained by HFH (see Section \ref{Sec:2D_repeated})
around this point shows two linear terms with opposite slopes $\Omega_1^2 = \pm 24.64 \kappa_x$ and two flat bands with slope close to zero that approximate well the exact curves.

\begin{figure}[h!]
\subfloat[Photonic crystal\label{fig:reconstruction2D_TM_PHC}]{\includegraphics[trim = 0mm 0mm 0mm 0mm, clip, width = 0.5\linewidth]{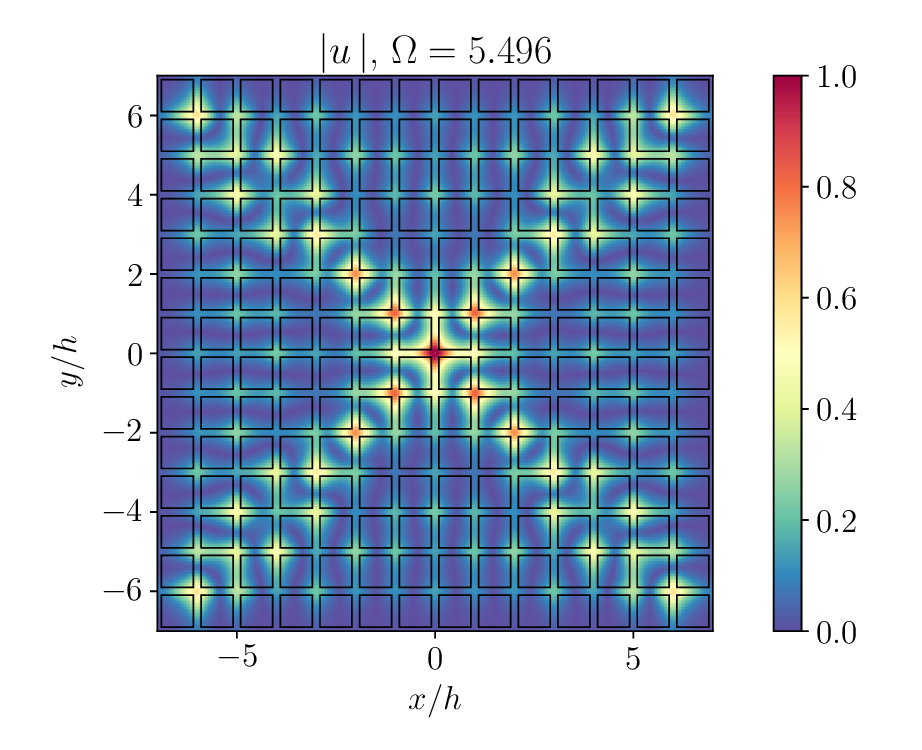}}
\subfloat[HFH\label{fig:reconstruction2D_TM_HFH}]{\includegraphics[trim = 0mm 0mm 0mm 0mm, clip, width = 0.5\linewidth]{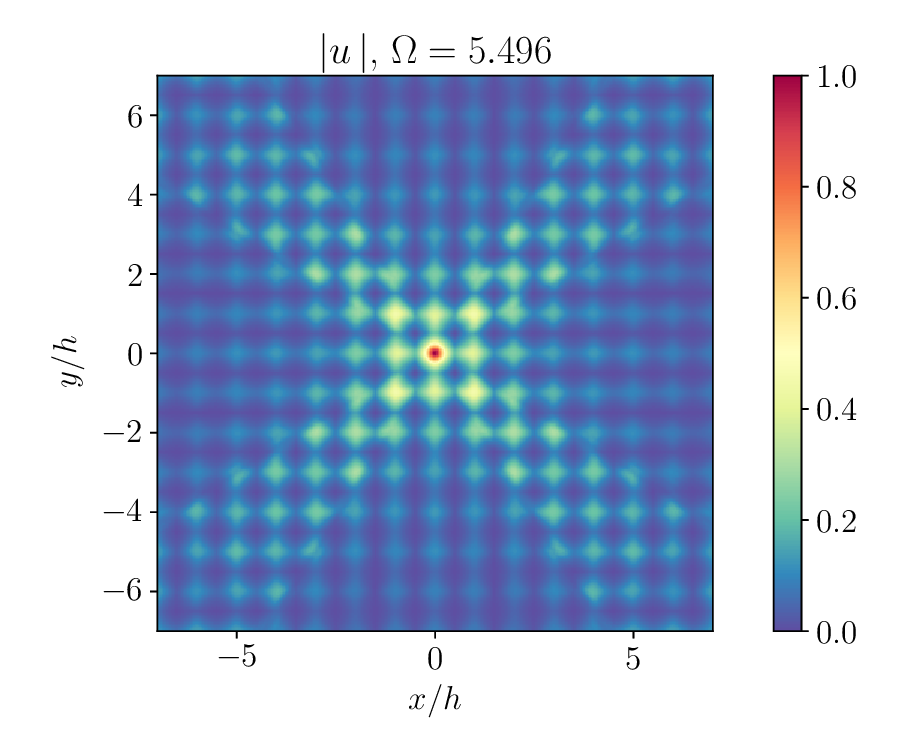}}
\caption{Field reconstruction for a finite square array of 196 square rods in TM polarization excited by a line source at the center with frequency $\Omega = 5.496$, close to the real part of an 
eigenfrequency of the periodic system $\Omega_0 =  5.442-0.008{\rm i}$ at point $\mathsf{X}$. 
The dynamic effective tensor \eqref{def_Tij} is $\mathbf{T}=\diag(-2.108 - 0.018{\rm i},\, 0.876 + 0.003{\rm i})$, indicating hyperbolic behaviour. In both figures the norm of the field has been normalized to its maximum value.}
\label{fig:reconstruction2D_TM}
\end{figure}

\begin{figure}[h!]
\subfloat[Photonic crystal\label{fig:reconstruction2D_TE_PHC}]{\includegraphics[trim = 0mm 0mm 0mm 0mm, clip, width = 0.5\linewidth]{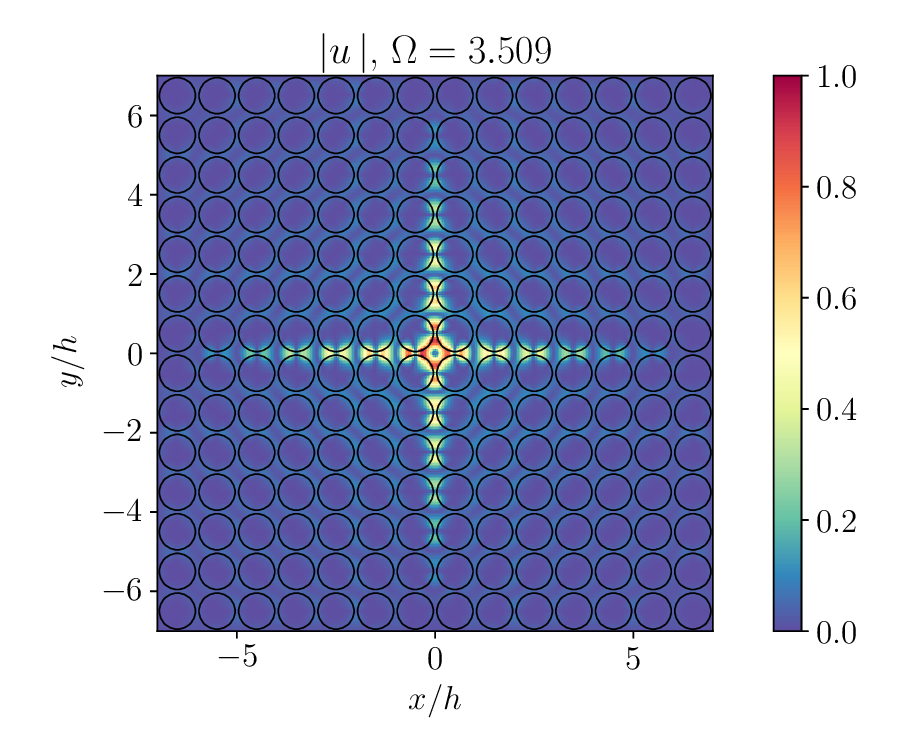}}
\subfloat[HFH\label{fig:reconstruction2D_TE_HFH}]{\includegraphics[trim = 0mm 0mm 0mm 0mm, clip, width = 0.5\linewidth]{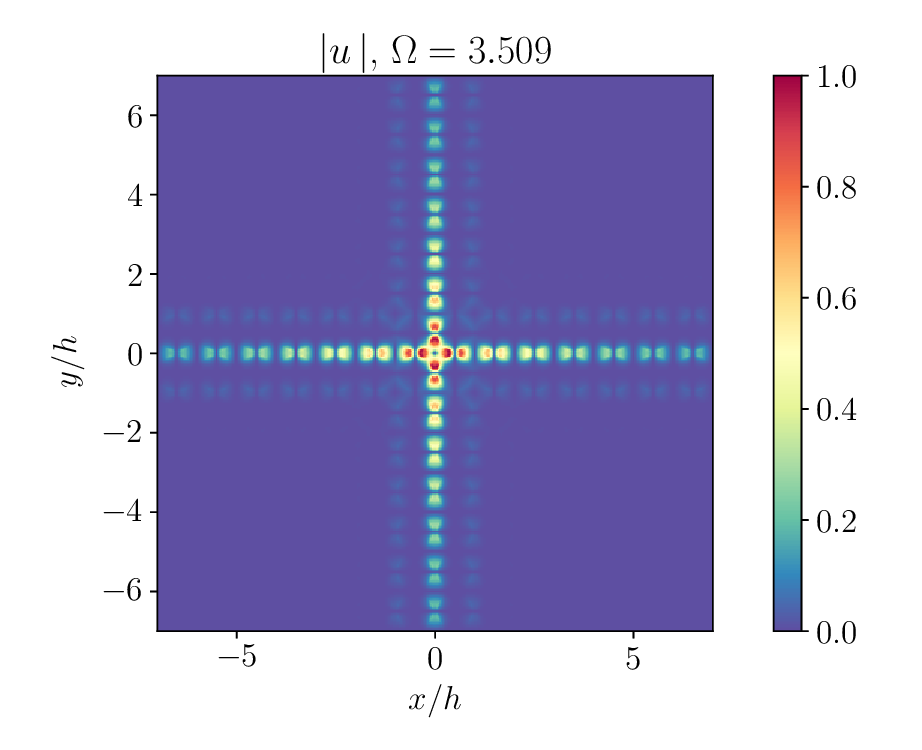}}
\caption{Field reconstruction for a finite square array of 196 circular rods in TE polarization excited by a line source at the center with frequency $\Omega = 3.509$, close to the real part of an 
eigenfrequency of the periodic system $\Omega_0 =  3.506-0.010{\rm i}$ at point $\mathsf{X}$. 
The dynamic effective tensor \eqref{def_Tij} is $\mathbf{T}=\diag(0.005 + 3\times 10^{-5}{\rm i},\, 0.950 - 0.012{\rm i})$, indicating parabolic behaviour. In both figures the norm of the field has been normalized to its maximum value.}
\label{fig:reconstruction2D_TE}
\end{figure}

\paragraph{Field approximation}
HFH theory has been developed in this paper for periodic infinite systems. However, in this paragraph, we use the asymptotic approximations obtained to describe finite-size systems, neglecting therefore the boundary effects: we now consider a 14 by 14 square array of rods with the same Drude permittivity as for the approximation of Figure \ref{fig:band_3_2d_phc_square} but with a smaller damping term $\gamma/2\pi= 0.005$. The finite photonic crystal is 
excited by a line source at the center with frequency close to the real part of an eigenfrequency of the periodic system near symmetry point $\mathsf{X}$. 
The first example is an array of square rods of size $L=0.806 h$ in TM polarization (see Figure \ref{fig:reconstruction2D_TM}). At the chosen frequency, 
the real parts of the coefficients $T_{ii}$ in the effective tensor of the HFH theory \eqref{def_Tij} of Section \ref{Sec:2D_simple} have opposite sign,
$\Re(T_{11}) = -2.108$ and $\Re(T_{22}) = 0.876$ and are of the same order of magnitude, leading to an effective hyperbolic behavior. The predicted theoretical wave field 
distribution is shown in Figure \ref{fig:reconstruction2D_TM_HFH}: the wave propagation is highly directive and is aligned along the
diagonals of the system. The resulting X-shape effect comes from the superposition of two effective media, for points $\mathsf{X}(\pi,0)$ and $\mathsf{M}(0,\pi)$ of the Brillouin zone. 
The full numerical simulations shown in Figure \ref{fig:reconstruction2D_TM_PHC} share this same qualitative feature.\\
Next we study the TE polarization case for cylindrical rods of diameter $D=0.91 h$. In this case, HFH predicts a distinct anisotropy aligned along the lattice axis, 
since $|\Re(T_{11})|\ll|\Re(T_{22})|$. This is indeed what we observe on full wave simulations shown in Figure \ref{fig:reconstruction2D_TE_PHC} with the same qualitative agreement 
for the solution of the effective parabolic equation shown in Figure \ref{fig:reconstruction2D_TE_HFH}. This directive emission due to the excitation of a surface plasmon-like mode, where the field is mostly confined at the interface between the dielectric background and the Drude metal, is well captured by the dispersive HFH theory, including the field decay as a result of material losses.

\section{Conclusion}
In this paper, we have extended the technique of high-frequency homogenization to dispersive periodic media, within which the physical properties do depend on the frequency. The work has been performed in both 1D and 2D and we considered different cases depending on the nature of the point around which we want to build an effective approximation. Near the edges of the Brillouin zone, we performed high-frequency homogenization for a frequency which is a simple eigenvalue, or a repeated one, and we also considered the case where several single eigenvalues are close to each other. Far from the edges, we developed an approximation in the case where no damping is considered. \\
In each of these cases, we were able to develop an approximation for both the dispersion diagram and the envelope function which defines the wavefield at the zeroth order. These asymptotic approximations come with an effective parameter or tensor that encapsulates the dispersive properties of the considered material. The results have been validated using comparisons with Finite Element Simulations for different configurations. We also discussed the interpretation of the effective parameter with respect to the nature of the wavefield.\\
Potential extensions of this work include pushing the asymptotics presented in this paper to higher orders, or extending it to
the case of waves in other periodic dispersive media: for example the full vector Maxwell's equations (such as in photonic crystal fibres within which TE and TM waves are usually fully coupled in oblique incidence), the Navier equations for fully coupled in-plane pressure and shear waves in phononic crystals, the Kirchhoff-Love equations for flexural waves in thin plates, or arrays of resonators (such as Helmholtz resonators, high-contrasted inclusions \cite{briane2014first,Bouchitte2015}, {or bubbles hosting Minnaert resonant frequencies \cite{ammari2019bloch}}).

\appendix 
\section{Expressions of $a$ and $b$ for the Lorentz model}\label{Sec:App1}
\paragraph{Physical parameters}
In the case of a Lorentz model, the physical parameters \eqref{setting_1d_a} and \eqref{setting_1d_b} are
\begin{equation}
\hat{a}_h(X,\omega)=\left\lbrace
    \begin{aligned}
   &{a}_0 &\text{ for } X\in(0,\phi h) \\
   &\hat{\mathbb{a}}(\omega)={a}_\infty \left( 1-\sum_{p\geq 0}\frac{\omega_{p,1}^2}{\omega(\omega+\mathrm{i}\hat{\gamma}_{p,1})-\omega_{D,p,1}^2}\right)^{-1} &\text{ for } X\in(\phi h,h) 
    \end{aligned}
    \right.
\end{equation}
and 
\begin{equation}
\hat{ b}_h(X,\omega)=\left\lbrace
    \begin{aligned}
   &{ b}_0 &\text{ for } X\in(0,\phi h) \\
   &\hat{\mathbb{b}}(\omega)={ b}_\infty \left( 1-\sum_{p\geq 0}\frac{\omega_{p,2}^2}{\omega(\omega+\mathrm{i}\hat{\gamma}_{p,2})-\omega_{D,p,2}^2}\right) &\text{ for } X\in(\phi h,h) 
    \end{aligned}
    \right.
\end{equation}
respectively, where $\omega_{p,i}$ stands for a plasmon frequency, $\hat{\gamma}_{p_i}$ for a damping coefficient, and $\omega_{D,p,i}$ for a Lorentz resonant frequency. When all the $\omega_{D,p,i}$ are zero, the Lorentz model becomes known as the Drude model. 
\paragraph{Adimensionalized parameters}
The adimensionalized physical parameters \eqref{a_adim_setting} and \eqref{rho_adim_setting} for this model then read 
\begin{equation}
\label{a_adim}
a(\xi,\Omega)=\left\lbrace
    \begin{aligned}
   &1 &\text{ for } \xi\in(0,\phi) \\
   &\mathbb{a} (\Omega)=\frac{{a}_\infty}{a_0} \left( 1-\sum_{p\geq 0}\frac{\Omega_{p,1}^2}{\Omega(\Omega+\mathrm{i}{\gamma}_{p,1})-\Omega_{D,p,1}^2}\right)^{-1} &\text{ for } \xi\in(\phi,1) 
    \end{aligned}
    \right.
\end{equation}
and
\begin{equation}
 \label{rho_adim}
 b(\xi,\Omega)=\left\lbrace
    \begin{aligned}
   &1&\text{ for } \xi\in(0,\phi ) \\
   &\mathbb{b} (\Omega)=\frac{{ b}_\infty}{ b_0} \left( 1-\sum_{p\geq 0}\frac{\Omega_{p,2}^2}{\Omega(\Omega+\mathrm{i}{\gamma}_{p,2})-\Omega_{D,p,2}^2}\right) &\text{ for } \xi\in(\phi,1) 
    \end{aligned}
    \right.
\end{equation}
respectively, with for $i=1,\,2$:
\begin{equation*}
 \label{adim_plasmon_freq_diss}
 \Omega_{p,i} = \frac{\omega_{p,i}h}{c_0}\text{, }\Omega_{D,p,i} = \frac{\omega_{D,p,i}h}{c_0} \text{ and } \gamma_{p,i}=\frac{\hat{\gamma}_{p,i}h}{c_0}.
\end{equation*}

\section{Finite element formulation}\label{Sec:App2}

Using a Lorentz model, we write $\mathbb{a}(\Omega) =  \frac{N_a(\Omega)}{D_a(\Omega)}$ and 
 $\mathbb{b}(\Omega) =  \frac{N_b(\Omega)}{D_b(\Omega)}$, where $N_a$, $D_a$, $N_b$, $D_b$ are polynomials of $\Omega$.
We have to solve the following eigenproblem
\begin{equation}
    \nabla\cdotp \left[a(\xi,\Omega) \nabla u(\xi)\right] + \Omega^2 b(\xi,\Omega) u(\xi) = 0.
    \label{evpnl}
\end{equation}

The weak formulation of the problem is derived by multiplying Eq.~(\ref{evpnl}) 
by the complex conjugate of a test function $v$ and integrating the first term by part on the 
unit cell $\mathcal{Y}$:
\begin{equation}
    -\int_\mathcal{Y} a(\Omega) \nabla u(\xi)\cdotp\nabla v^*(\xi) {\rm d}\xi
    + \Omega^2 \int_\mathcal{Y} b(\Omega) u(\xi) v^*(\xi) {\rm d}\xi
    = 0
    \label{evpnl_weak}
\end{equation}
where the boundary term vanishes because of the quasi-periodic boundary conditions. 
We define for:
$$ 
\mathbb{A}_0(u,v) = -\int_{\mathcal{Y}\setminus\mathcal{Y}_i}\nabla u(\xi)\cdotp\nabla v^*(\xi) {\rm d} \xi, \qquad
\mathbb{A}_1(u,v) = -\int_{\mathcal{Y}_i}\nabla u(\xi)\cdotp\nabla v^*(\xi) {\rm d} \xi $$
and
$$ \mathbb{B}_0 (u,v)= \int_{\mathcal{Y}\setminus\mathcal{Y}_i} u(\xi) v^*(\xi) {\rm d}\xi, \qquad
\mathbb{B}_1 (u,v)= \int_{\mathcal{Y}_i} u(\xi) v^*(\xi) {\rm d}\xi.$$
Plugging the expression for $a$ and $b$ in (\ref{evpnl_weak}) and rearranging we get:
\begin{align*}
   \mathcal{T}(\Omega,u,v)  = & D_a(\Omega)D_b(\Omega)\left[ \mathbb{A}_0(u,v) +\Omega^2 \mathbb{B}_0(u,v) \right]\\
    &+N_a(\Omega)D_b(\Omega) \mathbb{A}_1(u,v) \\   
    &+ \Omega^2  D_a(\Omega)N_b(\Omega)  \mathbb{B}_1(u,v) = 0
\end{align*}
which is a polynomial eigenvalue problem solved using the open source FEniCS finite element library \cite{alnaesFEniCSProjectVersion2015} interfaced with the SLEPc eigensolver \cite{hernandezSLEPcScalableFlexible2005,camposParallelKrylovSolvers2016}.

\section*{Acknowledgments} 
MT and RA would like to thank the Isaac Newton Institute for Mathematical Sciences, Cambridge, for support and hospitality during the programme Mathematical theory and applications of multiple wave scattering where work on this paper was undertaken. This work was supported by EPSRC grant no EP/R014604/. BV and RVC are supported by the H2020 FET-proactive Metamaterial Enabled Vibration Energy Harvesting (MetaVEH) project under Grant Agreement No. 952039. SG and RVC were funded by UK Research and Innovation (UKRI) under the UK government’s Horizon Europe funding guarantee [grant number 10033143]. 
The authors also thank Ping Sheng for fruitful discussions about the accumulation points.

\bibliographystyle{unsrt}  
\bibliography{references}  
\end{document}